\documentclass[iop]{emulateapj}

\shorttitle{SMA identification of SSA22-AzTEC1}
\shortauthors{Tamura et al.}

\begin{document}

\title{Submillimeter Array identification of the millimeter-selected galaxy SSA22-AzTEC1: a protoquasar in a protocluster?}

\author{
	Y. Tamura\altaffilmark{1,2}, 
	D. Iono\altaffilmark{1}, 
	D.~J. Wilner\altaffilmark{3}, 
	M. Kajisawa\altaffilmark{4,5},
	Y.~K. Uchimoto\altaffilmark{6}, 
	D.~M. Alexander\altaffilmark{7}, 
	A. Chung\altaffilmark{3,8},
	H. Ezawa\altaffilmark{9}, 
	B. Hatsukade\altaffilmark{1,6}, 
	T. Hayashino\altaffilmark{4}, 
	D.~H. Hughes\altaffilmark{10}, 
	T. Ichikawa\altaffilmark{4}, 
	S. Ikarashi\altaffilmark{6},
	R. Kawabe\altaffilmark{1,11}, 
	K. Kohno\altaffilmark{6,12}, 
	B.~D. Lehmer\altaffilmark{7,13,14}, 
	Y. Matsuda\altaffilmark{7}, 
	K. Nakanishi\altaffilmark{9}, 
	T. Takata\altaffilmark{9}, 
	G.~W. Wilson\altaffilmark{15}, 
	T. Yamada\altaffilmark{4}, and 
	M.~S. Yun\altaffilmark{15}
}

\altaffiltext{1}{Nobeyama Radio Observatory, National Astronomical Observatory of Japan, 
		Nobeyama, Minamimaki, Minamisaku, Nagano 384-1305, Japan; \email{yoichi.tamura@nao.ac.jp}}
\altaffiltext{2}{Depertment of Astronomy, The University of Tokyo, Hongo, Bunkyo-ku, Tokyo 113-0033, Japan}
\altaffiltext{3}{Harvard-Smithsonian Center for Astrophysics, 60 Garden Street, Cambridge, MA 02138}
\altaffiltext{4}{Astronomical Institute, Tohoku University, Aramaki, Aoba, Sendai, Miyagi 980-8578, Japan}
\altaffiltext{5}{Research Center for Space and Cosmic Evolution, Ehime University, Bunkyo-cho, Matsuyama, Ehime 790-8577, Japan}
\altaffiltext{6}{Institute of Astronomy, The University of Tokyo, Osawa, Mitaka, Tokyo 181-0015, Japan}
\altaffiltext{7}{Department of Physics, Durham University, Durham DH1 3LE, UK}
\altaffiltext{8}{Department of Astronomy, Yonsei University, Seoul 120-749, Korea}
\altaffiltext{9}{National Astronomical Observatory of Japan, Osawa, Mitaka, Tokyo 181-8588, Japan}
\altaffiltext{10}{Instituto Nacional de Astrofisica, Optica y Electronica, Aptdo.\ Postal 51 y 216, 72000 Puebla, Mexico}
\altaffiltext{11}{Department of Astronomy, School of Science, Graduate University for Advanced Studies, Osawa, Mitaka, Tokyo 181-8588, Japan} 
\altaffiltext{12}{Research Center for the Early Universe, School of Science, The University of Tokyo, Hongo, Bunkyo, Tokyo 113-0033, Japan}
\altaffiltext{13}{The Johns Hopkins University, Homewood Campus, Baltimore, MD 21218}
\altaffiltext{14}{NASA Goddard Space Flight Center, Code 662, Greenbelt, MD 20771}
\altaffiltext{15}{Department of Astronomy, University of Massachusetts, 710 North Pleasant Street, Amherst, MA 01003}

% ==========================================================

\begin{abstract}
We present results from Submillimeter Array (SMA) 860-$\micron$ sub-arcsec astrometry and multiwavelength observations of the brightest millimeter ($S_\mathrm{1.1mm} = 8.4$~mJy) source, SSA22-AzTEC1, found near the core of the SSA22 protocluster that is traced by Ly$\alpha$ emitting galaxies at $z=3.09$. We identify a 860-$\mu$m counterpart with a flux density of $S_\mathrm{860\mu m } = 12.2 \pm 2.3$~mJy and absolute positional accuracy that is better than $0\farcs 3$. At the SMA position, we find radio to mid-infrared counterparts, whilst no object is found in Subaru optical and near-infrared deep images at wavelengths $\le 1~\micron$ ($J > 25.4$ in AB, 2$\sigma$). The photometric redshift estimate, using flux densities at $\ge 24~\micron$, indicates $z_\mathrm{phot} = 3.19^{+0.26}_{-0.35}$, consistent with the protocluster redshift. We then model the near-to-mid-infrared spectral energy distribution (SED) of SSA22-AzTEC1, and find that the SED modeling requires a large extinction ($A_V \approx$~3.4~mag) of starlight from a stellar component with $M_\mathrm{star} \sim 10^{10.9} M_{\sun}$, assuming $z = 3.1$. Additionally, we find a significant X-ray counterpart with a very hard spectrum ($\Gamma_\mathrm{eff} = -0.34 ^{+0.57}_{-0.61}$), strongly suggesting that SSA22-AzTEC1 harbors a luminous AGN ($L_\mathrm{X} \approx 3 \times 10^{44}$~ergs~s$^{-1}$) behind a large hydrogen column ($N_\mathrm{H} \sim 10^{24}$~cm$^{-2}$). The AGN, however, is responsible for only $\sim 10\%$ of the bolometric luminosity of the host galaxy, and therefore the star-formation activity likely dominates the submillimeter emission. It is possible that SSA22-AzTEC1 is the first example of a protoquasar growing at the bottom of the gravitational potential underlying the SSA22 protocluster. 
\end{abstract}


% ==========================================================

\section{Introduction}

Mounting evidence
suggests that submillimeter-selected galaxies \citep[SMGs,][]{Smail97, Hughes98, Barger98, Blain02, KScott08, Perera08, Austermann09a, Austermann09b, KScott10} are the most massive, gas-rich systems at $z \sim$ 2--3 
with far-infrared (FIR) luminosity of $L_\mathrm{FIR} \sim 10^{12}$--10$^{13} L_\sun$ \citep[e.g.,][]{Borys05, Greve05, Solomon05, Tacconi06, Dye08}. Their extreme luminosity is likely produced by intense episodes of star-formation, and therefore it is likely that SMGs are undergoing rapid growths of their stellar components. A main catalyst for starburst is believed to be major mergers of gas-rich galaxies \citep[e.g.,][]{Smail98}. The merger event is probably an efficient mechanism to transfer cold gas into the nuclear region and fuel the star-formation and nuclear activity, as suggested by hydrodynamical simulations \citep[e.g.,][]{Hopkins06, Narayanan10}.

% - SMGs and BHs
An outstanding issue surrounding the SMG population is that dusty starburst galaxies at high redshift can relate to early growth of super-massive black holes seen at the present day. 
It is believed that radiation pressure from accretion disks around super-massive black holes efficiently strips away the obscuring gas and dust that fuel star-formation activities in their host galaxies, making accreting massive black holes an important regulator of galaxy formation and evolution \citep[e.g.,][]{Silk98, Fabian99, DiMatteo05}. Early (sub)millimeter surveys have revealed that host galaxies of powerful active galactic nuclei (AGNs) at high-$z$ are often submillimeter-bright \citep[e.g.,][]{McMahon94, Dunlop94, Isaak94, Ivison95, Archibald01, Stevens03, Reuland04, Stevens10}. 
Multiwavelength diagnostics indicate that a substantial fraction (20\%--50\%) of SMGs have AGNs at their center, suggesting a causal connection between the SMG phenomena (i.e., bulge formation) and massive black hole growth \citep{Alexander03, Alexander05c, Borys04, Swinbank04, Takata06, Menendez-Delmestre09, Hainline09}.
SMGs have a volume number density and a redshift distribution similar to those of quasars \citep[e.g.,][]{Chapman05}, allowing many to propose a hypothesis that the populations of SMGs and quasars can be evolutionarily linked. Moreover, black hole masses for SMGs appear to be systematically smaller than those found in quasars \citep{Alexander08}, implying that some fraction of SMGs with obscured X-ray sources are at a {\it protoquasar} phase, where the black hole is growing more rapidly than in typical galaxies and is about to blow off the thick surrounding gas clouds by accretion-related outflows from the black hole \citep[e.g.,][]{Sanders88, Kawakatu03, Kawakatu06, Kawakatu07, Granato04, Granato06, Alexander05b, Borys05, Kawakatu09}.

% A lot of studies have provided evidence for co-evolution of bulges and central black holes in galaxies in the local (REF) and distant Universe (REF).

% - accelerated growth of galaxies and BHs
Theoretical studies of cosmic structure formation in a cold dark matter (CDM) universe \citep[e.g.,][]{Kauffmann96} predict accelerated and correlated growths of galaxies and black holes in high density environments, such as protoclusters of galaxies, which are embedded within the most massive dark matter halos collapsing in the early Universe. 
 It is natural to expect that the merger rate should be enhanced in high-density regions of galaxies, and in fact, number excesses of SMGs have been tentatively claimed towards known overdensities around high-redshift radio galaxies \citep{Ivison00, Stevens03}.
Hence the idea  naturally arises that SMGs that host accreting massive black holes tightly relate to high-$z$ large-scale structures.

The $z = 3.09$ SSA22 protocluster was originally identified as a significant concentration, in redshift space, of Lyman break galaxies \citep{Steidel98}. The surface density of those Lyman break galaxies is $\sim$6 times higher than in the field, making this concentration one of the highest density regions know to date. Theoretical modeling indicates that the protocluster will evolve into a rich cluster with a total mass of $\gtrsim 10^{15} M_\sun$ at the present day \citep{Steidel98}. The protocluster has also been found to contain a factor of $\sim$3--6 overdensity of Ly$\alpha$ emitters \citep{Steidel00, Hayashino04}. Subsequent spectroscopic follow-up of these Ly$\alpha$ emitters, which are believed to be star-forming building blocks of galaxies, has revealed a large-scale filamentary structure extending across $\sim$60 $h^{-1}$~Mpc or more \citep{Matsuda05}. In this structure, many extended Ly$\alpha$-emitting objects \citep[Ly$\alpha$ blobs,][]{Steidel00, Hayashino04, Matsuda04} have been found, which are thought to be a formation site of massive galaxies \citep[e.g.,][]{Geach05, Geach07}. \citet{Uchimoto08} have suggested a density enhancement of another massive population of distant red galaxies \citep[DRGs,][]{Franx03, vanDokkum03}. Furthermore, \citet{Lehmer09a} have found a remarkable enhancement in AGN in SSA22 compared to $z \approx 3$ field galaxies. Thus, the SSA22 protocluster is an ideal site for studying the co-evolution of massive galaxies and powerful AGNs in a high-density environment.

Recently a large-area 1.1-mm survey toward SSA22 has been performed \citep{Tamura09} using the AzTEC 1.1-mm camera \citep{Wilson08a} mounted on the ASTE 10-m telescope \citep{Ezawa04}. They have mapped a $\approx$390~arcmin$^2$ region of the protocluster and revealed 30 SMGs. Among them, the brightest 1.1-mm source SSA22-AzTEC1 ($S_\mathrm{1.1mm}=8.4^{+0.8}_{-1.0}$~mJy) was found near the core of the protocluster, which hints that it could evolve into a massive elliptical in this descendant of the protocluster. 

However, the multiwavelength counterparts to SSA22-AzTEC1 cannot be easily identified because of the coarse beam (30$''$ in a full width at half maximum, FWHM) of the AzTEC/ASTE instrument and multiple candidates within the error circle. For SMG astrometry,
centimeter radio interferometers have often been employed so far, but they are not sensitive to SMGs at $z > 3$ since radio flux rapidly dims with increasing redshift. In contrast, imaging of the dust emission in submillimeter waveband benefits from the strong negative K-correction, and its flux density  is almost constant for a galaxy with a fixed FIR luminosity at redshifts $z \approx$ 1--10 \citep{Blain93}. 
Recent high spatial resolution observations \citep[e.g.,][]{Iono06a, Younger07, Younger09, Wang07, Wang09} using the Submillimeter Array \citep[SMA,][]{Ho04} have uncovered a substantial population of high-$z$ SMGs that are undetected in the radio and optical, suggesting that we would miss some high-$z$ SMGs by the radio selection method. Another potential hazard of using wavelengths other than (sub)millimeter is that a simple extraction of the nearest radio or mid-infrared (MIR) source within an error circle could lead to misidentification of counterparts to SMGs \citep[e.g.,][]{Cowie09} though it is difficult to estimate an accurate percentage of misidentifications. 
%\textbf{Therefore, a high-resolution submillimeter observation of SSA22-AzTEC1 is the inevitable next step in order to further investigate its properties.}

In this paper, we present the results from SMA 860-$\micron$ observations and multiwavelength properties of the brightest 1.1-mm source SSA22-AzTEC1 found in the SSA22 protocluster field. We use spectral energy distributions (SEDs) to constrain its photometric redshift. We also model an SED in the near-infrared (NIR) to MIR to estimate the physical properties of the stellar component. We then utilize X-ray data to identify AGN activity. The multiwavelength study of the 1.1-mm source provides a fairly unique opportunity to investigate the growth of cosmic hierarchical structures  over quite different spatial scales from AGNs ($\sim$1~pc) to large-scale structures ($\sim$10~Mpc): co-evolution of super-massive black hole and its host galaxy, and preferential growth of a massive galaxy with a powerful AGN in a high-density environment in the early Universe, which is one of the main concerns in recent astrophysics. 

% - cosmology
Throughout this paper, we will assume an $\Omega _\mathrm{M} = 0.3$, $\Omega _{\Lambda} = 0.7$ cosmology with $H_0 = 70$ km s$^{-1}$ Mpc$^{-1}$ as cosmological parameters. An angular scale of $1\arcsec$ corresponds to 7.6~kpc physical size at $z = 3.1$.

% ==========================================================

\section{Data and Observations}

\subsection{SMA Observations and Data Reduction}
The SMA is an interferometer operating at submillimeter wavebands (230, 350, and 690~GHz) that consists of eight 6-meter antennas located at Mauna Kea, Hawaii. We used the `Compact (C)' configuration of the SMA. Seven of the 8 antennas were operational in the observing track, which gave a range of projected baseline lengths of 6--60~meters, which corresponds to 7--90~k$\lambda$ (i.e., projected antenna separation lengths measured in units of an observing wavelength $\lambda$, which determine the resultant spatial resolution of interferometric images). The heterodyne receivers equipped with a superconductor-insulator-superconductor (SIS) mixer were tuned to 355.843~GHz (upper sideband, hereafter USB) and 345.843~GHz (lower sideband, hereafter LSB) for observing the continuum emission. We obtained 2~GHz total bandwidth in each sideband. The FWHM of the primary beam\footnote{The primary beam is a point response function of a single dish element of interferometers, and provides a field of view of an interferometric synthesized image.} at the frequency is 34$''$, which is comparable to the AzTEC/ASTE spatial resolution and enough to cover the $2\sigma$ error circle of the AzTEC/ASTE positional uncertainty ($\lesssim 20''$). 

The data were obtained on September 6, 2008. The conditions were `excellent' (zenith opacities at 225~GHz, $\tau _\mathrm{225GHz}$ = 0.04--0.06, and the typical root-mean-square (r.m.s.) phase fluctuation at 350~GHz, $\Delta \phi \simeq 10\degr$), and the double sideband system noise temperatures throughout the track were 150--300~K. The phase tracking center for the target source SSA22-AzTEC1 was set to R.A.\ (J2000) = $\mathrm{22^h17^m32 \fs 4}$, Decl.\ (J2000) = $+0\arcdeg 17' 35\farcs 47$, 
which is the centroid derived from AzTEC observations of SSA22-AzTEC1. Each visibility was integrated for 30~sec, which is short enough to avoid decorrelation caused by atmospheric phase fluctuations. The total integration time on SSA22-AzTEC1 was 8.8~ksec or 2.5~hours. The target source benefits from having a very good nearby gain calibrator, a 2-Jy radio-loud quasar 3C446 ($5.6\arcdeg$ away from the target). 3C446 was observed every 20 minutes to calibrate the gain variation. 3C454.3 and Uranus were observed at the end of the track for bandpass and flux calibration, respectively. The total integration time on 3C454.3 was 60~min. 

A baseline error of $\sim 0.1 \lambda$ will yield a systematic positional error of an order $0 \farcs 02$, which is much below the statistical error ($0\farcs 13$; see \S\ref{sect:results}). To further check our astrometry, a fainter, nearer radio-loud quasar J2218--035 (R.A.\ (J2000) = $\mathrm{22^h18^m52\fs 0377}$, Decl.\ (J2000) = $-3\arcdeg 35' 36\farcs 879$, which is 3.8$\arcdeg$ away from SSA22) with a precise known position was observed for verification of the accurate transfer of the phase solutions and checking the astrometric precision. 

The visibility data were calibrated using the \textsc{idl}-based SMA standard reduction package, \textsc{mir}. We did not need to flag out any of the data because of the excellent observing conditions. The flux density of a gain calibrator 3C446 at 860~$\mu$m was estimated to be $2.12 \pm 0.11$~Jy. The calibrated visibility data in USB and LSB were compiled and imaged (Fourier-transformed) using the \textsc{Miriad} \citep{Sault95} task, \texttt{invert}.  The resultant size of the natural-weighted synthesized beam is $3\farcs 43 \times 1\farcs 92$ (position angle P.A.\ = $34\fdg 3$).  Since the spatial frequency coverage is fairly poor because of short observing time and the declination very close to zero, the resultant map suffers from high side-lobe level of 60\%. We extensively CLEAN-ed the dirty image blindly (i.e., without constraining the region of interest to model the intensity field), restored the clean model, and convolved it with the clean beam (a 2-dimensional Gaussian with $3\farcs 43 \times 1\farcs 92$, P.A.\ = $34\fdg 3$). The uncertainty of absolute flux scaling is estimated to be better than 15\%.

% ==========================================================

\subsection{Multiwavelength Data}

We have multiwavelength data from the radio to the X-ray, which were already taken and cover the position of SSA22-AzTEC1. Here we provide a brief summary of the existing data and, in part, newly taken data.

We use 1.4~GHz (20~cm) radio data obtained towards SSA22 that were already published in \citet{Chapman03}. The SSA22 field was observed in the B configuration of the VLA, yielding a $\simeq 5''$ synthesized beam.  The on-source time $t_\mathrm{integ}$ = 12~hr resulted in an r.m.s.\ noise level over the AzTEC coverage to be $\simeq$ 12~$\mu$Jy~beam$^{-1}$. 

We retrieve MIR photometry data from the \textit{Spitzer} archive to search for a counterpart at MIR wavelengths and measure the flux densities. We use 3.6, 4.5, 5.8, 8.0~$\micron$ images obtained with the Infrared Array Camera \citep[IRAC,][]{Fazio04} and 24, 160~$\micron$ images obtained with the Multi-band Imaging Photometer for Spitzer \citep[MIPS,][]{Rieke04}. There is no 70~$\micron$ data at the position of SSA22-AzTEC1. The basic calibrated data (BCD) of IRAC and MIPS are processed through masking, flat fielding, background matching, and mosaicing using the \textsc{mopex} (Mosaicking and Point Source Extraction) software, which is a package developed at the Spitzer Science Center for astronomical image processing, along with calibrated data retrieved from the \textit{Spitzer} archive. 
Then point sources at 3.6, 4.5, 5.8, 8.0, and 24~$\micron$ are extracted using a \textsc{mopex} pipeline, \textsc{apex}, by fitting the point response function (PRF) for each band to each source candidate. 

The deep NIR imaging observations of the SSA22 region have been performed by \citet{Uchimoto08} using MOIRCS\footnote{Multi-Object InfraRed Camera and Spectrograph developed for the 8.2~m Subaru telescope. See also a web page at \url{http://www.naoj.org/Observing/Instruments/MOIRCS/} for more details.} \citep{Ichikawa06, Suzuki08} mounted on the Subaru telescope, Mauna Kea, Hawaii. A part of the NIR data are published elsewhere \citep{Uchimoto08}, but we append NIR data that are newly taken with MOIRCS/Subaru (Uchimoto et al.\ 2010, in preparation). Here we describe a brief summary of the MOIRCS observations. The total exposure times of the \textit{J, H, K$_S$}-band imaging towards SSA22-AzTEC1 were 5820, 2765, and 4541~sec, respectively. The stellar image sizes of the $JHK_S$ images are $0\farcs 4$--$0\farcs 5$, and the limiting magnitudes of the images are $J=25.42$, $H=24.99$, and $K_S = 25.08$~mag (2$\sigma$) in AB system in a $1\arcsec$ diameter aperture. 

% - photometry method in near/mid-IR
In the photometry, IRAC and MOIRCS flux densities were measured as follows. In order to achieve higher signal-to-noise ratio, we first smoothed the IRAC and MOIRCS images so that their PRFs match that of 8.0-$\micron$ image. Then we measured the flux densities of the NIR-MIR counterparts using a relatively small ($3\farcs 0$ in diameter) aperture to avoid source confusion. The aperture corrections were performed for all of the NIR-MIR data by multiplying a scaling factor. We estimated this scaling factor by comparing the $3\arcsec$-aperture flux and a total flux, both of which were obtained for the 8.0-$\micron$ image since the 8.0-$\micron$ image was less susceptible to confusion noise than the other (smoothed) images (see Figure~\ref{stamp}). The 8.0-$\micron$ total flux was measured with SExtractor \texttt{MAG\symbol{"5F}AUTO} \citep{Bertin96}.

Deep optical images in $B$, $V$, $R$, $i^{\prime}$, $z^{\prime}$ and NB497 bands taken with Suprime-Cam \citep{Miyazaki02} on Subaru is available at the position of SSA22-AzTEC1. NB497 is a narrow band filter centered at 497.7~nm, which is designed to search for strong Ly$\alpha$ emission in galaxies at $z$ = 3.06--3.12. The details of the observations are described in \citet{Hayashino04} and \citet{Matsuda04, Matsuda05}. The average profile of a point source in the final images has an FWHM of $1\farcs 0$. All of the optical photometry are measured with a $2\farcs 0$ diameter aperture, and are corrected for the Galactic reddening of $E(\bv) = 0.062$. The limiting magnitudes after the correction are 28.2, 27.9, 28.3, 28.4, 28.1, and 27.4 (AB, 1$\sigma$) in $B$, NB497, $V$, $R$, $i'$, and $z'$-band, respectively.

% - Chandra Deep Protocluster Survey
Recently, \citet{Lehmer09a, Lehmer09b} and \citet{Geach09} have reported the ultra-deep (400~ksec) \textit{Chandra}/ACIS-I observations  towards SSA22 (\textit{Chandra} Deep Protocluster Survey). The survey covers a solid angle of $\approx 330$~arcmin$^2$, and revealed 297 X-ray sources, which are listed in the ``Main Source Catalog''. The typical sensitivity limits were approximately $4.8 \times 10^{-17}$ and $2.7 \times 10^{-16}$ erg~cm$^{-2}$~s$^{-1}$ for the 0.5--2~keV and 2--8~keV bands, respectively. These limits correspond to rest-frame 2--8~keV and 8--32~keV luminosities of $3.7 \times 10^{42}$ and $2.1 \times 10^{43}$~ergs~s$^{-1}$, respectively.

% ==========================================================

\section{Results and Analyses}
\label{sect:results}

\subsection{SMA Results}
\label{sect:sma_results}

\begin{figure*}
	\includegraphics[scale=0.45]{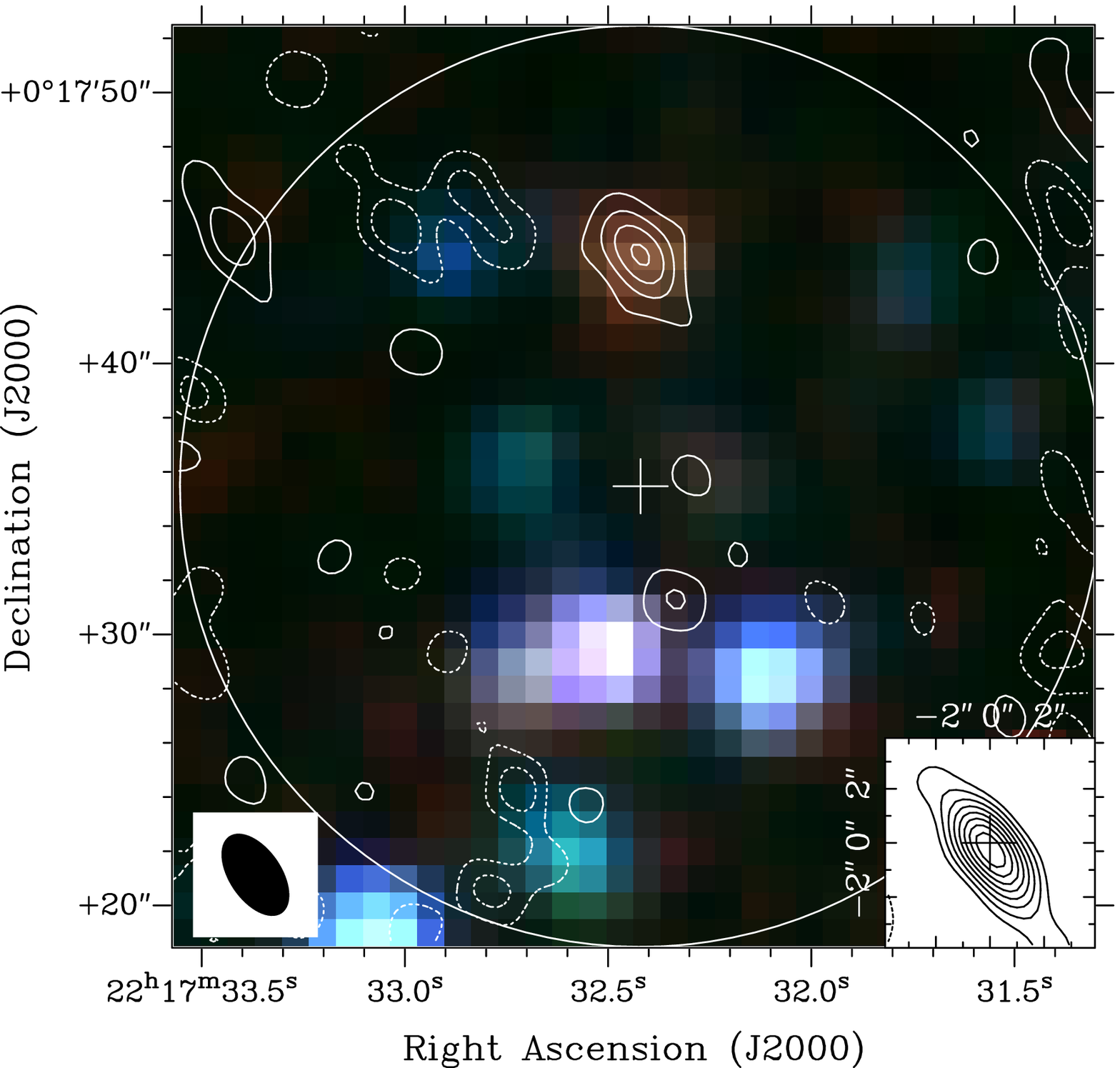}
	\includegraphics[scale=0.42]{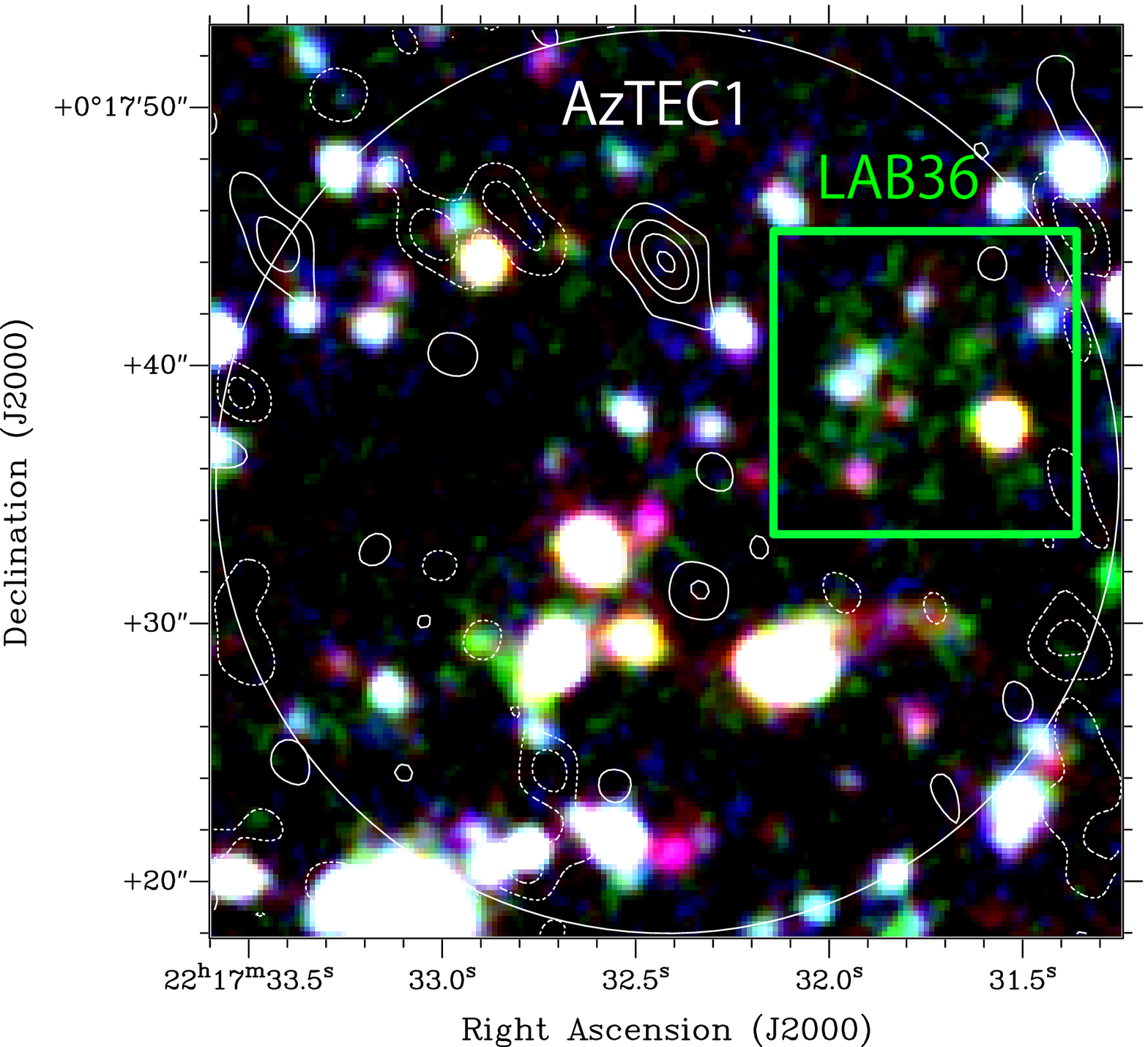}
	\caption{
			({\it left})
			The SMA 860~$\mu$m continuum image towards SSA22-AzTEC1 (contours), 
			overlaid on a \textit{Spitzer}/IRAC false color image (8.0, 4.5, 3.6~$\micron$ 
			for red, green, and blue, respectively). 
			The SMA image is natural-weighted and CLEANed. 
			The circle shows a field of view of the SMA (an FWHM of the primary beam, 
			which is the point response function of the SMA's single dish elements) 
			with a diameter of 34$''$. 
			The synthesized beam is indicated as an ellipse on the bottom-left corner. 
			Significant emission 8$''$ northward from the phase tracking center (cross) is evident. 
			The peak intensity is 11.8~mJy~beam$^{-1}$.
			The contours are drawn at (--4, --2, 2, 4, 6, 8)$\times \sigma$, 
			where $1\sigma$ = 1.4~mJy~beam$^{-1}$. 
			The negative levels are shown as dotted contours. 
			In the \textit{inset} of the left panel, we show
			the SMA 860-$\mu$m continuum image towards a test source 
			J2218--035. The systematic offset of the nominal peak position 
			from the phase tracking center (cross) is 
			($\Delta$R.A., $\Delta$Decl.) = ($0\farcs 00 \pm 0\farcs 13$, $-0\farcs 31 \pm 0\farcs 13$). 
			({\it right})
			The SMA 860~$\mu$m continuum image towards SSA22-AzTEC1, 
			overlaid on a Subaru/Suprime-Cam composite color image 
			(blue, green, and red for $B$-, NB497-, and $V$-band, respectively). 
			The green square denotes the position of the `mini' Ly$\alpha$ blob LAB36,
			which is located $\simeq 10''$ away from SSA22-AzTEC1. 
			The circle shows a field of view of the SMA (34$''$ in diameter).
			The contours are drawn as described above. 
			\label{fig1}}
\end{figure*}

In Figure~\ref{fig1}, we found a clear ($8.4\sigma$) emission feature (AzTEC~J221732.42+001744.0; we hereafter call the SMA source SSA22-AzTEC1 for convenience) located $\simeq 8''$ northward of the phase tracking center, strongly suggesting that this is the true counterpart to SSA22-AzTEC1. The resultant r.m.s.\ noise level is 1.4~mJy~beam$^{-1}$, which is derived from a region away from the 860 $\mu$m counterpart. The peak intensity at 860~$\mu$m is 11.8~mJy~beam$^{-1}$. The position is R.A.\ (J2000) = $\mathrm{22^h 17^m 32\fs 42}$, Decl.\ (J2000) = $+0\arcdeg 17'44\farcs 01$. 
The inset panel of Figure~\ref{fig1} (lower-right) shows the 860~$\mu$m continuum image obtained towards the test source J2218--035, a strong radio quasar with an accurately known position, which was clearly detected with $\simeq 10\sigma$. The offset of the nominal peak position from the phase tracking center is ($\Delta$R.A., $\Delta$Decl.) = ($0\farcs 00$, $-0\farcs 31$). The statistical positional error is estimated to be $0\farcs 13$ following the formula $\Delta \theta _\mathrm{stat} \simeq \sqrt{\theta_\mathrm{maj} \theta _\mathrm{min} }/ 2\, \mathrm{SNR}$, where $\theta_\mathrm{maj}$ and $\theta_\mathrm{min}$ are the FWHM along the major and minor axis of the synthesized beam, respectively, and $\mathrm{SNR}$ is the signal-to-noise ratio of the source in the cleaned map. These suggest that the phase transfer is applied correctly and the absolute positional uncertainty through this SMA observation is estimated to be $< 0\farcs 33$.

\begin{figure}
	\includegraphics[scale=0.32,angle=-90]{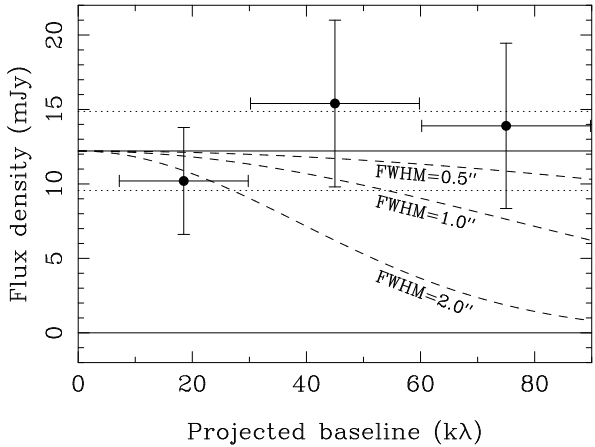}
	\caption{
		Visibility amplitudes versus projected baseline length for SSA22-AzTEC1 (solid circles).
		There is no evidence that SSA22-AzTEC1 is spatially resolved with the SMA beam. 
		The visibility amplitudes are well fitted to a constant value $12.2 \pm 2.7$~mJy (solid line)
		as a function of projected baseline length. The 1$\sigma$ confidence interval is shown as
		a region sandwiched between two dotted lines.
		The phase center is shift to the peak position of the 860-$\mu$m counterpart to SSA22-AzTEC1.
		We also show visibility amplitudes expected for axisymmetric 2-dimensional Gaussians with
		FWHM of $0\farcs 5$, $1\farcs 0$, and $2\farcs 0$.
		\label{fluxfit}}
\end{figure}

We then investigate the flux density and the spatial extent of the source more accurately in the $uv$-domain. Figure~\ref{fluxfit} shows the visibility amplitudes versus projected baseline length for SSA22-AzTEC1, along with those expected for axisymmetric 2-dimensional Gaussians with FWHM = $0\farcs 5$, $1\farcs 0$, and $2\farcs 0$. The amplitudes are well described by a constant value ($S_\mathrm{860\mu m} = 12.2 \pm 2.7$~mJy) as a function of projected baseline, suggesting that the source is not spatially resolved with the SMA $2''$ beam and the upper limit on the source size is $\lesssim 1''$. The flux densities estimated from the map and the visibilities are slightly different (and still consistent) even though the source is not spatially resolved. The main difference is that the image is synthesized from the vector sum of all measurements of the visibilities while the total flux estimate in the spatial frequency domain is a scalar estimate of the vector-averaged subsets of the visibility data, which is prone to noise-boosting by the fact that the visibility amplitude is always a positive quantity \citep[e.g.,][]{Thompson01}.

Considering the flux density at 1.1 mm is $S_\mathrm{1.1mm} = 8.4^{+0.8}_{-1.0}$ mJy, which is corrected for flux boosting, the 860-to-1100 $\mu$m flux density ratio is estimated to be $S_\mathrm{860\mu m}/S_\mathrm{1.1mm} = 1.45^{+0.47}_{-0.45}$. We scale our result to compare the flux ratio for 890~$\micron$ SMA follow-up of AzTEC sources, and find $S_\mathrm{890\mu m}/S_\mathrm{1.1mm} =1.4 \pm 0.4$ for SSA22-AzTEC1. This value is consistent with $1.6 \pm 0.7$ for the SMA-identified AzTEC sources reported by \citet{Younger07,Younger09}

% "massive" 'È‹â‰Í'ÌKƒoƒ"ƒh"™‹‰'́H
%    Schreiber+04: K<25.6 --> M*(median) = 0.8e11Mo --> should be seen in the Ks image at angular separation of 1-2"

% ==========================================================

\subsection{Possibility of Gravitational Lensing}
The intrinsic flux density of SSA22-AzTEC1 may be lower if there is an amplification by gravitational lensing. While recent theoretical works \citep[e.g.,][]{Paciga09} imply that the fraction of $\approx$10-mJy submm sources magnified by a factor of $> 2$ is relatively small ($\sim$5\%), there are many SMGs that are either known or strongly suspected to be lensed \citep[e.g.,][]{Ivison98, Downes03, Dunlop04, Kneib04, Motohara05, Wilson08b, Knudsen10, Swinbank10}. The small spatial size of SSA22-AzTEC1 measured with SMA (Figure~\ref{fluxfit}) suggests that the high brightness of SSA22-AzTEC1 is unlikely due to a strong gravitational lensing effect caused by a foreground massive galaxy that would happen to be aligned in the direct line of sight. 
Moreover, if not in the direct line of sight, a lensing object must be found at an angular separation comparable to the Einstein radius. If a source at $z \ge 1$ is located close to a foreground lensing object at $z \sim 0.3$, which is the favored geometry for efficiently lensing the background source, Einstein radii are $\sim 1''$--$2''$ and $\sim 30''$--$40''$ when the lensing object is a galaxy (dark halo mass of $10^{12}M_\sun$) and a cluster ($10^{15}M_\sun$), respectively.
 But, we do not see such a low-$z$ massive galaxy $\sim$$1''$--$2''$ away from SSA22-AzTEC1 although it would be easily detected in the Subaru $K_S$ image. 
In SSA22 we find neither X-ray clusters, which are often found at $z\sim 0.3$, nor higher-$z$ ($z\sim 1$) Sunyaev-Zel'dovich clusters, which should be observed in the AzTEC 1.1~mm map if exists. It is also unlikely that the SSA22 protocluster at $z=3.1$ itself is a lensing object because the SSA22 protocluster is by definition far from being virialized and should not have deep potential well similar to those of low-$z$ clusters. It is therefore likely that SSA22-AzTEC1 is not strongly magnified by gravitational lensing.

% ==========================================================

\subsection{Multiwavelength Counterparts}
\label{sect:multiw}

\begin{figure}
	\includegraphics[scale=0.78,angle=-90]{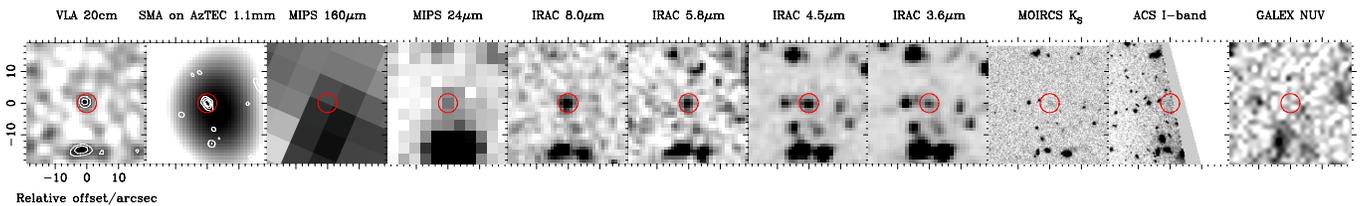}
	\caption{
		Postage stamp images of SSA22-AzTEC1 at radio 
		to near-ultraviolet wavelengths. 
		The red circles mark the SMA position. 
		\label{stamp}}
\end{figure}

At the position of the 860-$\mu$m source, we find a significant counterpart in the radio to MIR as well as in the X-ray, but the 860-$\micron$ source has no counterpart in the deep Subaru optical images. The flux densities and upper limits at radio to optical wavelengths are given in Table~\ref{tab1_Multiwavelength}. The postage stamp images at radio (20~cm) to near-ultraviolet (230~nm) wavelengths are presented in Figure~\ref{stamp}. 

% radio (VLA)
Although there are multiple counterpart candidates in the radio map, we identify a radio counterpart ($S_\mathrm{1.4GHz} = 42 \pm 12~\mu$Jy) at the position of the 860-$\mu$m source. 
The spectral index between 860~$\micron$ (350~GHz) and 20~cm (1.4~GHz) is $\alpha ^\mathrm{350}_\mathrm{1.4} = 1.03 \pm 0.15$, suggesting a redshift of $z \sim$~3--4 on the basis of the redshift-estimator proposed by \citet{Carilli99, Carilli00}. 

\begin{figure}
		\includegraphics[scale=0.4]{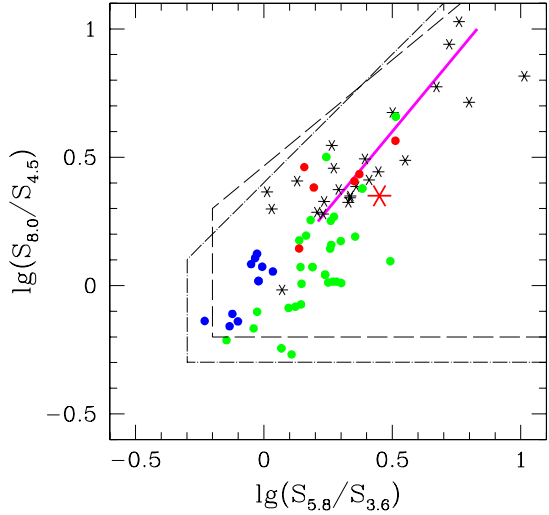}
		\caption{
			The {\it Spitzer}/IRAC $S_{5.8}/S_{3.6}$--$S_{8.0}/S_{4.5}$ color--color diagram for SMGs. 
			The filled circles in blue, green, and red are SMGs with millimeter CO or {\it Spitzer}/IRS 
			MIR spectroscopy at $z<1.5$, $1.5<z<3.0$, and $z>3.0$, respectively.  
			The big red star is SSA22-AzTEC1, and indeed it is found near the transition region between $z<3$ and $z>3$ SMGs. 
			The smaller, black stars represent infrared-luminous quasars from the FLS survey \citep{Lacy04}, 
			and it seems the $z>3$ SMGs indeed have colors that overlap with the infrared power-law AGNs (thick magenta line). 
			The color region on the plot that \citet{Yun08} proposed for the identification of SMG counterparts is outlined by dash-dotted line. 
			The region for infrared-luminous AGNs proposed by \citet{Lacy04} is outlined by the long dashed line. 
			\label{newcolor1}}
\end{figure}

% mid-IR (Spitzer)
We find counterparts to the SMA position in all IRAC bands and MIPS 24~$\mu$m. However, we see no significant counterpart at 160~$\micron$ mainly due to heavy blending from infrared sources, which are located $\sim 15\arcsec$ southward from the 860-$\micron$ position (see Figure~\ref{stamp}). 
These infrared sources have radio counterparts, and are the brightest at 24 $\micron$ in the SMA field of view. 
%
% - Min's diagnosis in IRAC bands
SSA22-AzTEC1 has red color in the {\it Spitzer}/IRAC bands, which is consistent with a $z \sim 3$ SMG. Figure~\ref{newcolor1} shows the color-color (IRAC 5.8-to-3.6 and 8.0-to-4.5 $\micron$ flux ratio) diagram for this source, SMGs with millimeter carbon monoxide (CO) or MIR {\it Spitzer}/IRS spectroscopy, and infrared-luminous quasars from the {\it Spitzer} First Look Survey \citep[FLS,][]{Lacy04}. The 5.8-to-3.6 and 8.0-to-4.5 $\micron$ flux ratios of this object are $\log{ (S_{5.8}/S_{3.6})} = 0.46^{+0.03}_{-0.04}$ and $\log{ (S_{8.0}/S_{4.5}) } = 0.35 ^{+0.02}_{-0.03}$, respectively. Indeed, it is found near the transition region between $z<3$ and $z>3$ SMGs with secure redshifts. It is seen in Figure~\ref{newcolor1} that infrared-luminous quasars identified in the FLS survey also have red IRAC colors, and it seems the $z>3$ SMGs indeed have colors that overlap with the infrared-luminous AGNs with a power-law spectrum in MIR (thick magenta line).  We note that SSA22-AzTEC1 has the IRAC colors consistent with a $z \sim 3$ SMG, but all $z>3$ SMGs also have IRAC colors very similar to the infrared power-law AGNs. 

% near-IR (MOIRCS)
The SED of SSA22-AzTEC1 is well constrained at wavelengths $\gtrsim 3~\micron$, but drops out at a wavelength of 1~$\micron$ ($J > 24.6$ in AB, 2$\sigma$) or shorter. 
We tentatively detect counterparts to SSA22-AzTEC1 in $H$- (3$\sigma$) and $K_S$-bands (4$\sigma$), whereas we see no emission in $J$-band. 
The observed color of SSA22-AzTEC1 in the NIR bands ($J-K > 1.0$) may meet the color criterion ($J-K > 1.4$ in AB) for distant red galaxies \citep[DRG,][]{Franx03, vanDokkum03}, which are often  found in the Universe at $z\sim 2$ and thought to be one of the most massive populations at this epoch \citep[e.g.,][]{Kajisawa06}.
This extremely red spectrum in the NIR--MIR strongly suggests that SSA22-AzTEC1 can have a large obscuring column in front of the stellar component and/or an AGN. 

% - optical-to-UV
We find no significant emission in the Subaru optical $B$, $V$, $R$, $i^{\prime}$, $z^{\prime}$ and NB497 images. The photometric data from the Subaru observations are tabulated in Table~\ref{tab1_Multiwavelength}. The non-detection in NB497 does not immediately imply that the SMA counterpart is not at $z \approx 3.1$ since, due to significant extinction, Ly$\alpha$ emission in $\sim 50\%$ of SMGs with a bright optical counterpart would not be strong enough to be detected in narrow-band imaging \citep[e.g.,][]{Chapman05}.
In addition, there are no counterparts in \textit{Hubble} optical (HST/ACS F814W) and GALEX near/far-ultraviolet images (see Figure~\ref{stamp}). The properties in the rest-frame optical images are very similar to those of submillimeter sources with secure (sub)millimeter interferometer identifications, GOODS~850-5 \citep[or GN10,][]{Pope05, Wang07, Wang09, Dannerbauer08, Daddi09a}, HDFN850.1 \citep{Hughes98, Downes99, Dunlop04, Cowie09}, SXDF850.6 \citep{Hatsukade10}, and SMA-identified AzTEC sources in the COSMOS field \citep{Younger07, Younger09}.

% - a Lyman-alpha Blob close to AzTEC1
It is worth noting that in the proximity of SSA22-AzTEC1 ($\simeq 10''$), we find a large emission nebula in the Subaru/NB497 image, which is a corroborated candidate for a Ly$\alpha$ blob at redshift $z=3.1$ \citep[LAB36,][]{Hayashino04}. In Figure~\ref{fig1}, we show a false-color optical ($B$, NB497, $V$) image around SSA22-AzTEC1. \citet{Hayashino04} have found 74 Ly$\alpha$ blobs over their surveyed area in SSA22 (699~arcmin$^2$), 59 of which are concentrated in the so-called high-density region (302~arcmin$^2$) of SSA22, 
in which SSA22-AzTEC1 is located.
 The probability that an object has a chance to be closely ($\le 10''$) associated with one of the 59 Ly$\alpha$ blobs over a 302~arcmin$^2$ region is 1.7\%. The separation angle of $\simeq$ 10$''$ corresponds to a projected separation of $\simeq 80 h_{70}^{-1}$~kpc in proper scale at $z = 3.1$. 

% - X-ray
We find a significant X-ray counterpart ($\approx$ 20 counts), which is listed as the source \#120 in the Main Source Catalog of the \textit{Chandra} Protocluster Survey, coincident with the SMA source position, suggesting the existence of an accreting massive black hole buried in the dust clouds. 
The position of the X-ray source is R.A.\ (J2000) = $\mathrm{22^h17^m 32 \fs 42}$ and Decl.\ (J2000) = $+0\arcdeg 17\arcmin 43\farcs 9$, and its associated positional error is $0 \farcs 54$, which is estimated from the 80\% confidence interval. The X-ray counterpart has a hard X-ray spectrum (the band ratio\footnote{Ratio of the count rates in the 2.0--8.0 and 0.5--2.0~keV bands.}, $BR =5.53^{+2.93}_{-2.29}$, or the effective photon index\footnote{A parameter for a power-law spectrum defined such that $f \propto \nu^{-\Gamma _\mathrm{eff}}$, where $f$ and $\nu$ are flux and frequency, respectively.}, $\Gamma _\mathrm{eff} = -0.34^{+0.57}_{-0.61}$) and is quite bright in the 2--8 keV band (but faint in the 0.5--2 keV band; see also Table~\ref{tab1_Multiwavelength}). In fact the X-ray spectrum is harder than that found for an AGN in a typical SMG among a sample of X-ray luminous SMGs at $z$ = 0.6--2.9 \citep{Alexander03, Alexander05c}, suggesting that it hosts one of the most heavily obscured distant AGNs known.  
We will further investigate the X-ray properties in \S~\ref{sect:smbh}.

% ==========================================================

\subsection{Photometric Redshift Estimates}
\label{sect:photoz}

The important next step is to estimate the redshift of SSA22-AzTEC1. It is extremely difficult to estimate its redshift through conventional optical or NIR spectroscopy because of the faintness in the optical and NIR. Here we estimate the photometric redshift (photo-$z$) by fitting the photometric data points to SED templates from stellar population synthesis models (\S~\ref{sect:NIR_photo-z}) and libraries that simulate MIR-to-radio emission from warm and cold interstellar medium (\S~\ref{sect:FIR_photo-z}).

% ==========================================================

\subsubsection{Photometric redshift from NIR data}
\label{sect:NIR_photo-z}

We use synthetic SED models of the composite stellar populations (CSP) with exponentially-declining star-formation activities from the \citet{Bruzual03} library. 
We use the models to fit the IRAC and MOIRCS photometry as well as the 2$\sigma$ upper limits in $i'$ and $z'$ bands listed in Table~\ref{tab1_Multiwavelength}, and compute the reduced-$\chi^2$ (i.e., $\chi^2/dof$, where $dof$ is the degree of freedom) as a function of redshift, 
\begin{eqnarray}\label{eqn:chi-square}
\chi ^2 = \frac{1}{N-p}\sum_{i}{\frac{ (S_{\nu, i}^\mathrm{obs} - S_{\nu}^\mathrm{model})^2}{\sigma_i^2}},
\end{eqnarray}
where $N$ is the number of photometric data points, $p$ is the number of free parameters, $S_{\nu, i}^\mathrm{obs}$ and $S_{\nu}^\mathrm{model}$ are the flux densities from observation and templates, respectively, and $\sigma_i$ is the uncertainty associated with each photometric datum.
Note that we do not use the photometry data at wavelengths $\lambda _\mathrm{obs} > 3(1+z)~\micron$ since emission from non-stellar components, such as hot dust, may contaminate the flux densities at $\lambda _\mathrm{rest} > 3~\micron$. We assume the \citet{Salpeter55} initial mass function and the solar metallicity ($Z = 1Z_{\sun}$).
We treat redshift, stellar mass, age, star-formation timescale\footnote{CSPs are modeled assuming exponentially declined star-formation activities. The star-formation timescale is defined as $e$-folding time-scale of the star-formation activities.}, and extinction as free parameters. We make use of the \citet{Calzetti00} extinction law to account for the extinction with color excess of $E(\bv) =$ 0.0--3.0, which corresponds to visual extinction of $A_V \approx$ 0--12.
We consider the redshift range from $z = 0$ to 7.

The photo-$z$ derived from fitting the stellar SED of SSA22-AzTEC1 is poorly constrained. We found a minimum of reduced-$\chi^2$ of 0.3 at $z_\mathrm{phot} \simeq 3.9$ with the confidence interval of $z \ge 1.0$ (99\% confidence level) or $z \ge 2.8$ (68\% confidence level). This means that the stellar population synthesis models at any redshift $z \ge 2.8$ can easily reproduce the MOIRCS and IRAC SED of SSA22-AzTEC1 and we virtually can not place strong constraints on the photometric redshift from fitting the MOIRCS and IRAC data to the stellar models. 

The primary reason for the poor constraint is that SSA22-AzTEC1 has no notable features in the rest-frame ultraviolet-to-NIR, such as the Lyman break, Balmer break, or the 1.6-$\micron$ stellar bump, which play critical roles in the ultraviolet-to-NIR photometric redshift estimates. As mentioned in the previous section, SSA22-AzTEC1 has an extremely red color in the rest-frame ultraviolet-to-NIR, suggesting a large extinction that makes a steeply declining SED towards shorter wavelengths. Such a steeply declining SED could cause errors in photometric redshift.
%: photo-$z$ codes can confuse these SEDs as Balmer breaks or other features, which can bias redshift toward higher redshift than true redshift. 

An example of such an error can be found in a series of papers on an SMG, GOODS~850-5 (or GN10). Wang et al. (2007, 2009) have reported extremely faint SED of this object over the rest-frame ultraviolet-to-NIR bands (an extremely deep limit to $K_S$(AB, 2$\sigma$) $< 27.2$ mag or 46 nJy). They also found that the photo-$z$ estimates for the ultraviolet-to-NIR data show $z_\mathrm{phot} \sim 6.5$, whereas the photo-$z$ from its FIR-to-radio ($24~\micron$--20~cm) SED shows $z_\mathrm{phot} \sim 4$. Eventually, \citet{Daddi09b} blindly searched for CO in this object and found the spectroscopic redshift of $z_\mathrm{CO} = 4.042$, which as a result rejected the photo-$z$ estimated from the rest-frame ultraviolet-to-NIR.

% ==========================================================

\subsubsection{Photometric redshift from MIR to radio data}
\label{sect:FIR_photo-z}

Another way to constrain the redshift is to use the FIR-to-radio SED. The ratio between MIR-to-(sub)mm and radio flux densities depends on the redshift. This is because in the local Universe the radio luminosity is tightly correlated with the FIR luminosity \citep[FIR-to-radio correlation,][]{Helou85, Condon92, Helou93, Yun01} and this relation seems to hold at high redshifts \citep[e.g.,][]{Garrett02, Appleton04, Ibar08, Seymour09, Ivison10a, Ivison10b}. There have been a number of studies exploiting this relation to constrain the photometric redshift of SMGs \citep[e.g.,][]{Carilli99, Carilli00, Hughes02, Aretxaga03, Aretxaga05, Aretxaga07, Wang09, Daddi09a}, and this method can be efficient when a galaxy suffers from extremely heavy extinction in the rest-optical/NIR bands.

We use the available photometry at 24, 860, 1100~$\micron$, and 20~cm data, and employ the SED library from \citet{Michalowski09} to estimate the redshift of SSA22-AzTEC1. 
% - model
The library consists of SED models, which were developed using the GRASIL code \citep{Silva98}, that were being fit to the photometric data points \citep[e.g.,][]{Hainline09} of 76 SMGs with spectroscopic redshifts \citep{Chapman05}. The mean FIR luminosity, stellar mass, and redshift, averaged over the 76 SMGs, are $L_\mathrm{FIR} = 10^{12.7} L_{\sun}$, $M_\mathrm{star} = 10^{11.7} M_{\odot}$, and $z=2.0$, respectively. To eliminate poorly constrained SEDs from the library, we chose SEDs with robust 24-$\micron$ photometry (i.e., SEDs that are well-constrained at MIR wavelengths and the Wien tail of the FIR dust emission). We do not use the SEDs that strongly underpredict flux densities at MIR or submillimeter wavelengths ($\lambda_\mathrm{obs} =$ 3.6--8.0, 850 $\micron$). For our final SED templates, we employ 35 SEDs and an averaged SED that is computed by averaging the 35 SEDs. The mean FIR luminosity, stellar mass, and redshift of the 35 SMGs are  $L_\mathrm{FIR} = 10^{12.8} L_{\sun}$, $M_\mathrm{star} = 10^{11.6} M_{\odot}$, and $z=2.0$, respectively, which are consistent with those averaged over the 76 spectroscopically-identified SMGs.
% - method
Following Equation~\ref{eqn:chi-square}, we calculate the $\chi^2$ values between the photometric data of SSA22-AzTEC1 and each SED model with varying redshift and luminosity as free parameters. The $\chi ^2$ fit to the data at the observed wavelengths of $\lambda _\mathrm{obs} \ge 24~\micron$ provides a good fit, but the $\chi ^2$ fit is poor when we include the $\lambda _\mathrm{obs} \le 8~\micron$ data. The $\lambda _\mathrm{obs} \le 8~\micron$ data likely suffer from heavy extinction by dust if the data at $\lambda _\mathrm{obs} \le 8~\micron$ represent the stellar component of SSA22-AzTEC1. Furthermore the SED models are biased against an optically faint population of SMGs as noted by \citet{Michalowski09}. Therefore we use the photometric data at $\lambda _\mathrm{obs} \ge 24~\micron$ hereafter. 

\begin{figure}
	\includegraphics[scale=1.0]{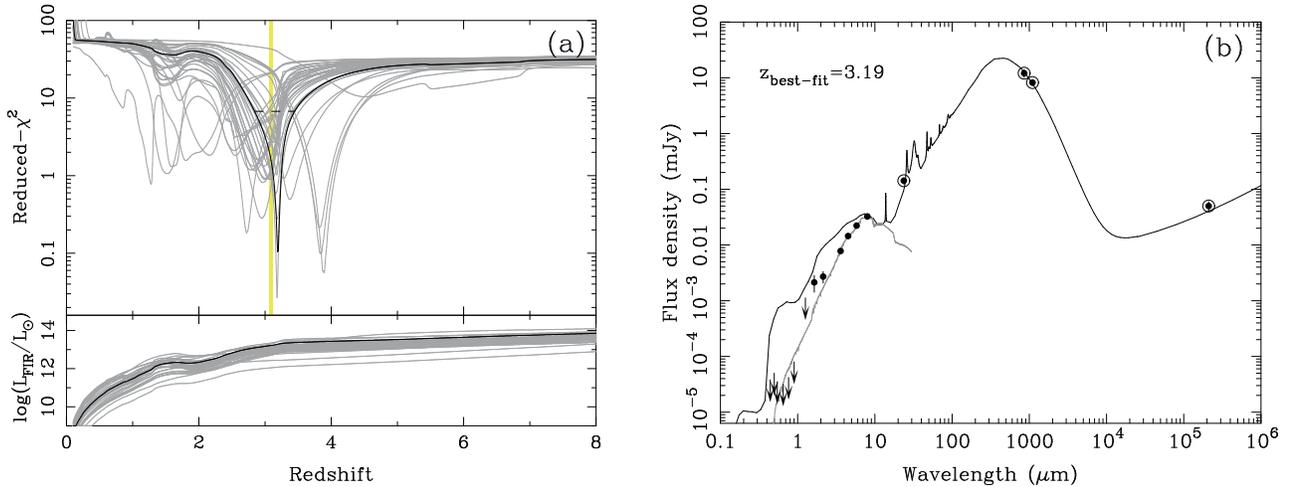}
	\caption{
		Radio-to-MIR photometric redshift of SSA22-AzTEC1. 
		(a) The top panel shows the reduced-$\chi^2$ versus redshift. 
		Template SEDs are from \citet{Michalowski09}. 
		The grey curves show the reduced-$\chi^2$ for 35 SEDs fit to 24-$\micron$-detected 
		SMGs with spectroscopic redshift. 
		The thick solid curve shows that for an SED averaged over the 35 SEDs, and
		the horizontal dashed line shows the 99\% confidence interval. 
		The yellow bar represents the redshift range of the protocluster ($z$ = 3.06--3.12). 
		The bulk of the curves have their local minima at the protocluster redshift, 
		and hence we adopt $z = 3.1$ in the rest of the analyses. 
		The bottom panel shows the infrared luminosities that allow the minimum $\chi^2$ value at each redshift. 
		The details are described in \S~\ref{sect:FIR_photo-z}.
		(b) The observed flux densities (filled circles) and the SED \citep[curve,][]{Michalowski09} 
		averaged over 35 SMGs with spectroscopic redshift \citep{Chapman05}. 
		The SED is shifted at the best-fit redshift ($z = 3.19$) and is scaled to 
		the best-fit FIR luminosity ($\log{L_\mathrm{FIR}}=13.3$). 
		Only the data at $\lambda _\mathrm{obs} \ge 24~\micron$ are used in the $\chi^2$-fit (open circles). 
		The thin curve indicates the best-fit SED model \citep{Maraston05} 
		for the stellar component of SSA22-AzTEC1. 
		The rapid dimming and break at $\lambda_\mathrm{obs} \le 8~\micron$ is likely 
		due to obscured stellar (and possibly AGN) emission that suffers from 
		extremely heavy attenuation ($A_V \approx 3.4$) of gas and dust in SSA22-AzTEC1. 
		See the text (\S~\ref{sect:ir_sed}) for more details.
		\label{chi2fit}}
\end{figure}

% - result
Figure~\ref{chi2fit}a (top) shows the reduced $\chi^2$ values for the 35 SED models with 24-$\micron$ data (thin curves) and a model averaged over the 35 SEDs (a thick curve), as a function of redshift. Figure~\ref{chi2fit}a (bottom) shows the best-fit FIR luminosity that gives the least $\chi^2$ value at each redshift. The $\chi ^2$ of the bulk of the SEDs have their local minima at $z \approx 3$. The $\chi ^2$ minima, however, scatter broadly in redshift space probably because we have only two degrees of freedom and many of the SEDs presented in \citet{Michalowski09} remain poorly constrained especially in the MIR band. 

%\textbf{A significant excess in the rest-frame MIR band in SED templates, which is likely attributed to emission from hot dust heated by an AGN, may systematically  raise the best-fit redshift. An MIR excess can raise FIR luminosity compared to templates without an obvious MIR excess, and this may tend to also elevates the radio luminosity following the radio-to-FIR correlation. To explain the observed flux densities at wavelengths where positive K-correction works (i.e., MIR and radio), the SED templates with MIR excess have to be more redshifted than those without MIR excess.}

AGN-dominant galaxies could have high dust temperatures, which increases the MIR luminosities and shifts the FIR peak of SEDs to shorter wavelengths. Thus redshift estimates using such SEDs may tend to be biased toward higher-$z$. In fact, there are just three SED templates\footnote{SMM J105238.30+572435.8, SMM J123600.15+621047.2, SMM J163650.43+405734.5} having their local minima at $z \approx 3.8$.  Independent measurements of their spectra at rest-frame ultraviolet to MIR wavelengths \citep{Swinbank04, Takata06, Menendez-Delmestre07, Menendez-Delmestre09, Pope08, Hainline09} have suggested that they are probably AGN-dominant (i.e., AGN is responsible for $\gtrsim$50\% of the bolometric luminosity of its host). The AGN of SSA22-AzTEC1 does not appear to dominate the host bolometric luminosity as discussed in the following section (\S~\ref{sect:smbh}), suggesting that the fits to the AGN-dominated SEDs might not be appropriate. 
On the contrary, SED templates with lower dust temperature can decrease the best-fit redshift. In our SED fits, local minima are evident at $z \sim 1.5$. The 9.7-$\micron$ silicate absorption feature falls to the 24-$\micron$ data point at $z \simeq 1.5$, slightly improving the $\chi^2$-fit for this particular redshift. However, the $\chi^2$ values at local minima around low redshifts $z \sim 1.5$ are much larger than those for SEDs with best-fit redshift of $z \sim$ 3--4, and the stellar SED fits do not favor redshifts $z \sim 1$. This suggests that the redshift of $z \sim 1.5$ is unlikely.

% - averaged SED
It is thus reasonable to focus on the averaged SED. The photometric redshift from the averaged SED is $z_\mathrm{phot} = 3.19^{+0.26}_{-0.35}$ (the error bar is estimated from the 99\% confidence interval), consistent with the redshift of the protocluster ($z$ = 3.06--3.12). The derived FIR luminosity is $L_\mathrm{FIR} = 1.9^{+0.4}_{-0.6} \times 10^{13} L_{\sun}$ (the error bar is estimated from the 99\% confidence interval). The large FIR luminosity implies a star-formation rate of $\sim 4 \times 10^3 M_{\sun}$~yr$^{-1}$ based on \citet{Kennicutt98}. The best-fit averaged SED along with the photometric data points is shown in Figure~\ref{chi2fit}b. We note that if we take the best-fit SED model (SMM~J123711.98+621325.7) that has the least minimum-$\chi^2$ among all of the SEDs, the inferred photometric redshift is $z = 3.18^{+0.26}_{-0.41}$, again consistent with the protocluster redshift. 
We will hence assume $z = 3.1$ for the redshift of SSA22-AzTEC1 in the rest of the analyses.

% ==============================================

\subsection{NIR-to-MIR Spectral Energy Distribution}
\label{sect:ir_sed}

% - break at < 5 micron
While the \citet{Michalowski09} SED model is in good agreement with the data at observed wavelengths of $\lambda _\mathrm{obs} \gtrsim 24~\micron$, we see a significant dimming/break in the SED at $\lambda _\mathrm{obs} \lesssim 8~\micron$ compared with the SED model. This is probably because of (i) extremely heavy dust extinction of the stellar component, (ii) low-mass stellar component, and/or (iii) a steep power-law spectrum in rest-frame NIR bands due to hot dust heated by a buried AGN.
% (iv) a stellar component populated by a top-heavy \citep{Baugh05} or bottom-light \citep{vanDokkum08} initial mass function. 

% - fit to population synthesis models
To further investigate the possibilities (i) and (ii) shown above, we use stellar population SED models to fit the MOIRCS and IRAC photometric data. The method used here is the same as introduced in \S~\ref{sect:NIR_photo-z}, but we use population synthesis models from the \citet{Maraston05} and \citet{Bruzual03} libraries and fix the redshift to $z = 3.1$. 
%To further investigate the possibilities (i) and (ii) shown above, we use stellar population SED models of the composite stellar populations (CSP) with exponentially-declined star-formation activities from the \citet{Maraston05} and \citet{Bruzual03} libraries. In the SED fit, we apply the MOIRCS and IRAC flux densities listed in Table~\ref{tab1_Multiwavelength}. We consider the NIR-MIR photometric data under the assumption that all of the NIR-to-MIR emission are attributed to stars (not an AGN). We assume the redshift of $z = 3.1$, the \citet{Salpeter55} initial mass function and the solar metallicity ($Z = 1Z_{\sun}$). We also make use of the \citet{Calzetti00} extinction law to account for the red, faint spectrum in NIR-to-MIR bands. 

The best-fit SED models from \citet{Maraston05} and \citet{Bruzual03} are essentially consistent, and each of the best-fit SED suggests 
$A_V = 3.4\pm0.2$~mag, age of $2.5^{+3.0}_{-1.0} \times 10^{7}$~yr, and stellar mass of $M_\mathrm{star} = 7.3^{+7}_{-1.7} \times 10^{10} M_{\sun}$ ($\chi ^2 = 0.963$, from the Maraston model) and $A_V = 3.4^{+0.4}_{-0.3}$~mag, age of $2.5^{+5.6}_{-0.9} \times 10^{7}$~yr, and $M_\mathrm{star} = 8.2^{+9}_{-1.5} \times 10^{10} M_{\sun}$ ($\chi ^2 = 0.886$, from the Bruzual \& Charlot model). The uncertainties represent the 1$\sigma$ confidence interval. 
We model CSPs with $e$-folding time-scale of star-formation activities ranging $\tau _\mathrm{SF}$ = 0.1--20~Gyr, but we cannot place a good restriction on the time-scale $\tau_\mathrm{SF}$ because the best-fit SED model is fairly young compared to $\tau _\mathrm{SF}$. We note that the ages from the stellar model fits are known to be somewhat unreliable, especially for stellar populations in the early Universe \citep{Shapley05}.
The extinction is comparable to but slightly larger than those found in SMGs with rest-frame optical spectra obtained by \citet[][$A_V = 3.0\pm 1.0$]{Swinbank04}, \citet[][$A_V = 1.7\pm 0.2$]{Borys05}, \citet[][$A_V = 2.9\pm 0.5$]{Takata06}, and \citet[][median $A_V$ = 2.03 $\pm$ 0.95]{Michalowski09}. \citet{Borys05} have studied stellar components of 13 SMGs found in the GOODS-N field, and found the stellar masses of $\sim 10^{11}$--$10^{12} M_{\sun}$ and a mean stellar mass of $\sim 2 \times 10^{11} M_\sun$.
The stellar mass of SSA22-AzTEC1 is slightly smaller than those for a typical SMG in GOODS-N. But it is larger than that of coeval ultraviolet-selected star-forming galaxies reported by \citet[][$\log{M_\mathrm{star}/M_\sun} = 10.32 \pm 0.51$ for $z\sim 2$]{Shapley05} and \citet[][$\log{M_\mathrm{star}/M_\sun} = 10.62 \pm 0.11$ for $z\sim 3$]{Magdis10}, suggesting that SSA22-AzTEC1 is more massive than a coeval typical star-forming galaxy. %, but might still undergo its initial build-up stage.
The SED fit improves ($\chi ^2 = 0.106$) if we employ a super-solar metallicity $Z = 2.5 Z_{\sun}$ although the best-fit parameters do not change drastically ($A_V = 2.4^{+0.5}_{-0.3}$, age of $1.6^{+8.4}_{-0.6} \times 10^7$~yr, and $M_\mathrm{star} = 3.3^{+8}_{-0.5} \times 10^{10} M_{\sun}$). Such higher metallicities are often favored in modeling SEDs of local ultra-luminous infrared galaxies \citep[e.g.,][]{Farrah05}. We note, however, that since the number of the photometric data points at $\le$~8~$\micron$ is limited to just six, the parameters (i.e., $A_V$, age, and $M_\mathrm{star}$) that determine the stellar SED are highly degenerate and could have large systematic uncertainties.

% ==============================================

\subsection{Luminous AGN Buried Deeply in SSA22-AzTEC1}
\label{sect:smbh}
Here we investigate the detailed properties of the X-ray counterpart referred in \S~\ref{sect:multiw}.
%
% - modeling the X-ray spectrum -- Nh and Lx
We utilize the \textsc{pimms} package%
\footnote{
\textsc{pimms} (Portable, Interactive Multi-Mission Simulator) is maintained by Koji Mukai, and is available at \url{http://heasarc.nasa.gov/docs/software/tools/}. 
}%
 to model the X-ray spectrum that accounts for the small effective photon index of $\Gamma _\mathrm{eff} = -0.34^{+0.57}_{-0.61}$. We simulate the intrinsic power-law spectra with photon indices $\Gamma$ = 1.4--2.0 behind hydrogen columns in the range $N_\mathrm{H} = 1 \times 10^{23}$--$4 \times 10^{24}$~cm$^{-2}$. The spectra are redshifted to $z=3.1$ and adjusted for absorption through the Galactic hydrogen column towards SSA22-AzTEC1 \citep[$4.1\times 10^{20}$~cm$^{-2}$,][]{Kalberla05}. Other components such as reflection and scattering are not considered. The band ratios between the soft (0.5--2~keV) and hard band (2--8~keV), $BR$, are estimated using \textsc{pimms} (v.~3.7a) appropriate for the \textit{Chandra}/ACIS-I Cycle 8 observations. For $BR > 3$, band ratios do not substantially depend on intrinsic photon index, but depend largely on hydrogen column density.
If the AGN has intrinsically a power-law spectrum with a photon index of $\Gamma = 1.8$, which is a typical value for nearby AGNs \citep{Tozzi06}, the hard spectrum of SSA22-AzTEC1 strongly suggests that the hydrogen column has $N_\mathrm{H} \sim 1 \times 10^{24}$~cm$^{-2}$ and is almost Compton-thick. 
It should be noted that  the band ratio works well for AGNs with $N_\mathrm{H} \lesssim 5 \times 10^{23}$~cm$^{-2}$ in determining the hydrogen column density, but it becomes less reliable for higher column densities because other components such as reflection and scattering make the X-ray spectrum more complex than just an absorbed power law. However, the X-ray spectral slope is as flat as the heavily-obscured SMGs reported by \citet{Alexander05c} which also appear to have $N_\mathrm{H} \sim 10^{24}$~cm$^{-2}$, based on more complex X-ray spectral fitting. 

Assuming that the X-ray source is associated with SSA22-AzTEC1 and it is located at $z = 3.1$, the absorption-corrected luminosity in rest-frame 0.5--8~keV, $L_\mathrm{X}$, is estimated following the expression,
\begin{eqnarray}
L_\mathrm{X} = 4 \pi d_\mathrm{L}^2 F_\mathrm{X}^\mathrm{obs} (1+z)^{\Gamma -2},
\end{eqnarray}
where $d_\mathrm{L}$ = 26.4~Gpc is the luminosity distance at $z=3.1$ and $F_\mathrm{X}^\mathrm{obs}$ is the absorption-corrected flux in the observed-frame 0.5--8 keV. We find $L_\mathrm{X} \approx 3 \times 10^{44}$~ergs~s$^{-1}$, which is slightly larger than those found in $z \sim 2$ AGN-classified SMGs \citep[$L_\mathrm{X} \sim 10^{44}$~ergs~s$^{-1}$, see Figure~8 in][]{Alexander05c}. It is therefore likely that SSA22-AzTEC1 hosts a heavily obscured AGN, which is one of the most luminous objects found among the SMG population. 

% - X-to-FIR luminosity ratio --- What powers AzTEC1?
Given the large X-ray luminosity, the bolometric output from the central AGN in SSA22-AzTEC1 can heat up the interstellar dust and it may contribute to a substantial fraction of the FIR luminosity of the host galaxy. It is becoming clear that AGNs are likely present in a significant fraction of SMGs but their contribution to the FIR emission is in many cases minor \citep[e.g.,][]{Alexander05c, Menendez-Delmestre09, Hainline09, Michalowski09}. Nevertheless, it is important to investigate the origin of the FIR luminosity of SSA22-AzTEC1 (and hence the intrinsic bolometric output from the AGN) because the AGN-to-host luminosity ratio can be used as an indicator of AGN activity and may provide a hint to the evolutionary phase of the central massive black hole. We now explore what powers the bolometric luminosity of SSA22-AzTEC1---massive star-formation or an AGN. 

The unabsorbed X-ray to FIR luminosity ratio of SSA22-AzTEC1 is $L_\mathrm{X}/L_\mathrm{FIR} = 0.004$. This value is very similar to that found by \citet{Alexander05c} (median $L_\mathrm{X}/L_\mathrm{FIR} \approx 0.004$), but it is an order of magnitude smaller than those found in quasars \citep[$L_\mathrm{X}/L_\mathrm{FIR} \approx 0.05$,][]{Elvis94, Alexander05c}. The reason why the luminosity ratio in SSA22-AzTEC1 is so much smaller than those in quasars is the extreme FIR luminosity of SSA22-AzTEC1.
%
% - an explanation for the derivation process of the bolometric correction and luminosity.
In order to estimate the bolometric luminosity of the AGN, we introduce the bolometric correction, which is defined as $f_\mathrm{bol} \equiv L_\mathrm{bol}^\mathrm{AGN}/L_\mathrm{X}$, where $L_\mathrm{bol}^\mathrm{AGN}$ is the bolometric luminosity of the AGN. The bolometric correction for the X-ray band is known to be $\sim$10--100 \citep[e.g.,][]{Elvis94} and to have a moderate dependence on the bolometric AGN luminosity. \citet{Marconi04} have investigated AGNs in the local Universe, and found an empirical relation between $L_\mathrm{bol}^\mathrm{AGN}$ and $f_\mathrm{bol}$ in 2--10~keV band as
\begin{eqnarray}\label{eq:bollum}
	\log{f_\mathrm{bol}} = 1.54 + 0.24\mathcal{L} + 0.012\mathcal{L}^2  - 0.0015 \mathcal{L}^3,
\end{eqnarray}
where $\mathcal{L}$ is a bolometric luminosity measured in units of $10^{12} L_\sun$ and is defined as $\mathcal{L} \equiv \log{L_\mathrm{bol}^\mathrm{AGN}}-12 = \log{f_\mathrm{bol}} + \log{L_\mathrm{X}^\mathrm{2-10keV} -12}$.
We correct the SSA22-AzTEC1's 0.5--8~keV luminosity to 2--10~keV, and obtain $L_\mathrm{X}^\mathrm{2-10keV} = 10^{10.7} L_{\sun}$ assuming $\Gamma = 1.8$. We solve Equation~(\ref{eq:bollum}) in terms of $\mathcal{L}$ by substituting $\log{f_\mathrm{bol}} = \mathcal{L}-\log{L_\mathrm{X}^\mathrm{2-10keV}}+12 = \mathcal{L}+1.3$ and find $\mathcal{L} = 0.33$. We then have $L_\mathrm{bol}^\mathrm{AGN} \approx 10^{12.3} L_\sun$ and $f_\mathrm{bol} \sim 40$ for the AGN of SSA22-AzTEC1. The bolometric correction is consistent with those measured for high-$z$ AGNs \citep[$f_\mathrm{bol}^\mathrm{2-10keV} \approx 35$,][]{Elvis94}. 

This unabsorbed bolometric luminosity of the AGN is extremely large, and as luminous as the most luminous objects in the local Universe \citep[e.g., Mrk~231,][]{Soifer87}. Nonetheless, the bolometric AGN luminosity can only account for $\sim$10\% of the huge bolometric output of SSA22-AzTEC1 ($L_\mathrm{FIR} \approx 2 \times 10^{13} L_\sun$). If we adopt the highest estimate for a bolometric correction of the X-ray luminous SMGs ($f_\mathrm{bol}^\mathrm{2-10keV} \approx 70$), the percentage will increase by a factor of $\sim$1.6. This suggests that the large FIR luminosity of SSA22-AzTEC1 is not mainly powered by the AGN, but should be attributed to other power sources, likely massive star-formation activities.

% ==========================================================

%% - Discussions 

\section{Discussions}

We have shown that SSA22-AzTEC1 is a hyper-luminous starburst galaxy at high redshift, likely $z = 3.1$, and harbors a luminous X-ray source. Here we discuss a possible scenario for the growths of the stellar component and massive black hole of SSA22-AzTEC1 in the protocluster environment.

% ==========================================================

\subsection{A Progenitor of a Massive Galaxy in a Protocluster}
\label{sect:massiveE}

The inferred redshift and the large stellar mass shown above suggest that we are possibly witnessing the violent forming stage of a massive early-type galaxy that nucleates at the bottom of the gravitational potential underlying {\em this protocluster}, which can be the archetypical formation site of massive galaxies predicted from the current standard model of structure formation in a CDM universe. 
There is a claim that the bright-end of the evolved systems selected at NIR wavelengths is well populated by $z \sim 2$ but much less in $z \sim 3$ protoclusters around powerful radio galaxies, implying that the bright-end of the evolved, likely massive ($\sim$$10^{11} M_\sun$) galaxies had first appeared between $z$ = 2--3 \citep{Kodama07}. The time duration from $z=3$ to 2 is $\approx$1.1~Gyr whereas just 0.01~Gyr (i.e., comparable to a lifetime of a massive star) is required for SSA22-AzTEC1 to have $10^{11} M_\sun$ of stars if the stars could be produced at current star-formation rate ($\sim 4 \times 10^3 M_\sun$~yr$^{-1}$) and add another $\sim$(2--3)$\times 10^{10} M_\sun$ to this system. 
Although we have no information on the mass of the gas reservoir in SSA22-AzTEC1, it is possible for this object to evolve into a $M_\mathrm{star} \sim 10^{11} M_\sun$ system by $z=2$. A search for a molecular reservoir in this object will allow us to assess this possibility.

% A lot of studies have provided evidence for co-evolution of bulges and central black holes in galaxies in the local (REF) and distant Universe (REF).
% the luminous X-ray source found in this object suggests the existence of a rapidly-growing massive black hole at its center, and thus the bulge can also build up rapidly, which help back up

% The bulk of $M_\mathrm{star} \sim 10^{11} M_{\sun} galaxies at the present day are actually classified as ellipticals, many of which are often seen in local rich clusters (REF). Early-type galaxies dominates the 10^{11} M_{\sun} galaxies even at higher redshifts of $z = 0.4$--1 (REF).

% ======================================

\subsection{Growth of a Massive Black Hole in a Protocluster}
\label{sect:protoquasar}

% - Black hole mass
Comparing the black hole and stellar mass is a good method for understanding the evolutionary status of the AGN in SSA22-AzTEC1. While estimating masses of black holes ($M_\mathrm{BH}$) is often fairly difficult, there are ways to constrain the black hole mass. One of the most plausible ways using the available data for SSA22-AzTEC1 is to assume the Eddington ratio, $\eta$ (i.e., ratio of the bolometric to Eddington luminosity of the AGN). When $\eta = 1$, the AGN is powered by Eddington-limited accretion onto the black hole, which virtually provides a lower limit on an estimate of the black hole mass. 

 \citet{Alexander08} have investigated $M_\mathrm{BH}$ and $\eta$ of spectroscopically-identified SMGs in the \textit{Chandra} Deep Field-North. They divided these SMGs into two samples: one is a sample which consists of SMGs that exhibit optical broad emission lines. The other consists of SMGs that host X-ray identified obscured AGNs with no apparent broad-line region, which is more representative of the overall SMG population than the broad-line SMG sample. The average redshifts of both samples are consistent with the median redshift obtained for radio-identified SMGs \citep[$\langle z \rangle _\mathrm{median} =2.2$,][]{Chapman05}, and their stellar masses are well constrained by \citet{Borys05}. 
 For the broad-line SMGs, one can directly estimate $M_\mathrm{BH}$ and $\eta$ using the well-established virial black-hole mass estimator \citep[e.g.,][]{Wandel99, Kaspi00}. 
 %
 % - a summary for results from Alexander+08 
They found the mass ratios $M_\mathrm{BH}/M_\mathrm{GAL}$ of the broad-line SMGs and the X-ray-obscured SMGs to be $\sim 2 \times 10^{-3}$ and $\approx 2.9 \times 10^{-4}$, respectively. The average mass ratio of the broad-line SMGs is higher than that of the X-ray-obscured SMGs but lower than those of typical unobscured quasars at $z \sim 2$ \citep[$M_\mathrm{BH}/M_\mathrm{GAL} \sim$(5--10)$\times 10^{-3}$,][]{Peng06, Coppin08}. This suggests that the broad-line SMGs have more massive, rapidly growing black holes than the X-ray-obscured SMGs, and they are possibly at a transition phase between X-ray-obscured (i.e., typical) SMGs and unobscured quasars. 

% - AzTEC1's BH mass and ratio
Although uncertainties are considerably large, the inferred mass of the black hole in SSA22-AzTEC1 is $M_\mathrm{BH,Edd} \sim 5 \times 10^{7} M_\sun$ for Eddington-limited accretion, or $M_\mathrm{BH} \sim 2.5 \times 10^8\, (\eta/0.2)^{-1} M_\sun$.
The black-hole--to--galaxy mass ratio of SSA22-AzTEC1, which is estimated from the black hole mass divided by the stellar mass, is $M_\mathrm{BH}/M_\mathrm{GAL} = 3.5 \times 10^{-3}(\eta/0.2)^{-1}$. The mass ratio is (1.4--3.5)$\times 10^{-3}$ if the Eddington ratio has a realistic value between $\eta$ = 0.2--0.5, which are found to be true in the SMG population \citep{Alexander08}. 
This mass ratio is $\sim$5--10 times higher than those found for the X-ray-obscured SMGs, but it is below the values found in the $z\sim 2$ unobscured quasars. This large discrepancy might be attributed to the underestimation of the stellar mass, which is difficult to estimate as discussed in \S~\ref{sect:ir_sed}. If we underestimate the stellar mass by as large as a factor of 2 \citep[e.g.,][]{Shapley05}, the mass ratio could still be 2.5--5 times larger than those of the X-ray-obscured SMGs. A more likely explanation is that SSA22-AzTEC1 has a more massive black hole accreting at a higher rate (i.e., higher $\eta$) than the X-ray-obscured SMGs, but it is less massive than black holes of the quasars. In contrast, the mass ratio is comparable to those of the broad-line SMGs, suggesting that the black hole in SSA22-AzTEC1 can be in the same evolutionary stage as those in the broad-line SMGs. 
Although the constraints are very uncertain, it is hence possible that the black hole in SSA22-AzTEC1 is being at a transition phase to a quasar, or a {\it protoquasar} phase \citep[e.g.,][]{Sanders88, Kawakatu03, Kawakatu06, Kawakatu07, Granato04, Granato06, Alexander05b, Kawakatu09}, where the black hole is growing more rapidly than in typical SMGs and is still buried deeply in the central gaseous, dusty `cocoon'. 

% - Is AzTEC1 a protoquasar residing in a protocluster of galaxies at high redshift?
What is the importance of  finding a luminous accreting massive black hole, which is one of the most obscured AGN found in SSA22 and likely resides at the heart of the brightest SMG in the SSA22 protocluster? 
Large-area submillimeter surveys have suggested that SMGs are related to large-scale structures at high redshift \citep{Blain04, Scott06, Daddi09a, Daddi09b, Tamura09, Viero09, Weiss09b} even though there is a claim that the population of SMGs cannot always be used as tracers of the most massive dark halos at high-$z$ considering the short time duration (as short as 10~Myr) of their starburst phase \citep{Chapman09, Aravena09}. In particular, \citet{Tamura09} have found an angular correlation between the positions of the bright SMGs and the Ly$\alpha$ emitters in SSA22, strongly suggesting the physical association of the bright SMGs with the protocluster. 
 In addition, \citet{Almaini03} have reported strong clustering signals between SMGs and \textit{Chandra} X-ray sources. 
 These observational facts suggest that SMGs which host accreting massive black holes can tightly relate to large-scale structure at high-$z$, but the evidence for SMGs with heavily-obscured AGNs (i.e., protoquasars) forming in high-$z$ protoclusters has not reported. %This is probably because of the rareness of protoquasars, which is likely attributed to a low volume density of SMGs and a short duration of the pre-quasar phase, and observational difficulties to find such objects.

\begin{figure}[htbp]
	\includegraphics[scale=1.0]{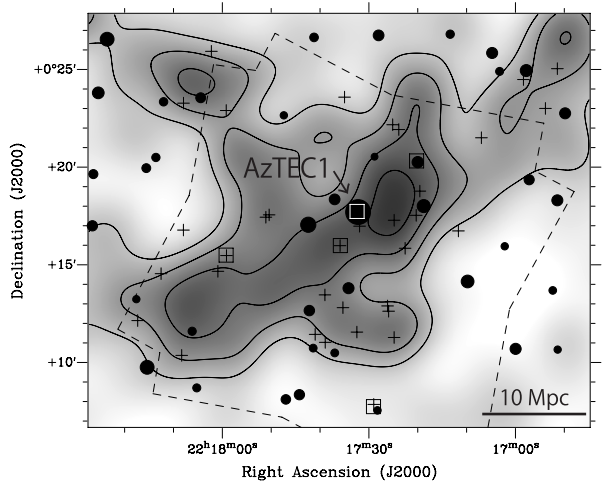}
	\caption{
		The position of SSA22-AzTEC1 and the large-scale distributions of luminous objects towards/in the SSA22 protocluster; AzTEC sources \citep[filled circles, whose sizes are proportional to 1.1-mm flux densities;][]{Tamura09}, robust candidates for $z=3.1$ Ly$\alpha$ blobs \citep[crosses;][]{Matsuda04}, and the most luminous $z=3.1$ AGNs in SSA22 with rest 0.5--8~keV luminosities of $L_\mathrm{X} \approx$ (2--3)$\times 10^{44}$~ergs~s$^{-1}$ \citep[squares;][]{Lehmer09a, Geach09}. The white square denotes the heavily-obscured AGN of SSA22-AzTEC1. The gray-scales show the projected number density of Ly$\alpha$ emitters at $z=3.1$ \citep{Hayashino04} and the contours are drawn at (1, 1.5, 2.0, 2.5) times the mean density of the Ly$\alpha$ emitters, which outlines the $z=3.1$ protocluster. The dashed line shows the field-of-view of the \textit{Chandra} Deep Protocluster Survey. The bar at bottom-right corner indicates a comoving scale of 10~Mpc at redshift of $z=3.1$.
	}\label{Environment}
\end{figure}

% - positions of AzTEC1 in the SSA22 protocluster
Figure~\ref{Environment} shows the position of SSA22-AzTEC1 and the large-scale distributions of luminous objects towards/in the SSA22 protocluster. The high-density region of Ly$\alpha$ emitters in SSA22 consists of at least three large-scale filamentary structures, and the highest density peak (i.e., the apparent protocluster core) actually corresponds to the intersection of these large-scale filaments \citep{Matsuda05}. SSA22-AzTEC1 is located close to the intersection of the filamentary structures, suggesting that this submillimeter-luminous protoquasar can trace the deepest central potential of the large-scale structure, possibly becoming a powerful AGN, just like high-$z$ radio galaxies and quasars in protoclusters \citep[e.g.,][]{McMahon94, Dunlop94, Isaak94, Ivison95, Archibald01, Stevens03, Reuland04, Stevens10}. Although the accurate redshift of SSA22-AzTEC1 is yet to be determined, this may be the first observational example supporting the long-standing prediction that powerful AGNs preferentially form in high-density environments such as protoclusters \citep[e.g.,][]{Sijacki07, Colberg08}. 

Finally, we compare SSA22-AzTEC1 with other object with AGNs found in SSA22. 
Within the protocluster, at least over the \textit{Chandra} coverage, no $z=3.1$ AGNs that exceeds SSA22-AzTEC's AGN in X-ray luminosity have been reported. As shown in Figure~\ref{Environment}, however, there are three X-ray-identified AGNs at $z=3.1$ in the innermost part of the protocluster, whose rest-frame 0.5--8~keV luminosities \citep[$L_\mathrm{X} \approx$ (2--3)$\times 10^{44}$~ergs~s$^{-1}$,][]{Lehmer09a, Geach09} are comparable to that of SSA22-AzTEC1. They have been identified in a Ly$\alpha$ emitter and two Ly$\alpha$ blobs in SSA22. 

An $S_\mathrm{1.1mm}\approx 4$-mJy source \citep[SSA22-AzTEC6,][]{Tamura09} could be associated with the X-ray luminous Ly$\alpha$ emitter even though there is no submillimeter interferometric identification. This implies that it might be another example of a protoquasar in the protocluster. However, the effective photon index of the X-ray source ($\Gamma _\mathrm{eff} = 1.68^{+0.21}_{-0.20}$) is consistent with that of an unobscured AGN, and the host galaxy is clearly visible in the optical broad-band images unlike SSA22-AzTEC1. These suggest that gas surrounding the AGN and stellar component is becoming to clear up. 
In contrast, the two Ly$\alpha$ blobs within the high-density region are faint at 1.1-mm\footnote{A SCUBA 3.8$\sigma$ detection has been reported for one of the Ly$\alpha$ blobs \citep[LAB14,][]{Geach05}, but we find no significant emission ($S_\mathrm{1.1mm} < 2$~mJy) in the better-quality AzTEC map.}, which does not match the prediction for hosts of protoquasars that vigorous star-formation is associated with the early growth of massive black holes. The X-ray sources of the Ly$\alpha$ blobs are unobscured ($\Gamma _\mathrm{eff} > 1$), which may imply the absence of dense gas reservoir surrounding the AGNs. 
Many studies predict that a luminous and unobscured phase of AGN evolution can follow the optically-thick phase at which the host is luminous in the FIR and the AGN is buried in dust clouds, as mentioned above \citep[e.g.,][]{Kawakatu03, Kawakatu06}. 
Therefore, we speculate that the AGNs in the Ly$\alpha$ emitter and blobs are possibly at more advanced phase than SSA22-AzTEC1's AGN.
%Further discussions of these AGNs in the Ly$\alpha$ galaxies are beyond the scope of this paper. 

% - What is a protoquasar?
% Many theoretical works (Archibald+02, Granato+04) have predicted that
% Emergency of QSOs is delayed relative to spheroid formation
% the growth of black hole mass is delayed relative to the spheroid

% ==========================================================

\section{Conclusions and Future Prospects}

Using the Submillimeter Array, we have identified an 860-$\micron$ counterpart to an $S_\mathrm{1.1mm} = 8.4$-mJy source, SSA22-AzTEC1, which has been previously discovered in a 1.1-mm single-dish survey towards the SSA22 protocluster at $z=3.1$. We then have investigated intrinsic nature of this source using multiwavelength data all the way from the radio to the X-ray. The derived properties of SSA22-AzTEC1 is summarized in Table~\ref{tab2_Summary}. Our main findings and conclusions are as follows. 

\medskip 

1. We identified an 860-$\mu$m counterpart with a flux density of $S_\mathrm{860\mu m } = 12.2 \pm 2.3$~mJy and absolute positional accuracy that is better than $0\farcs 3$. The SMA counterpart is not resolved with the SMA 2$\arcsec$ beam. The high brightness of this object is not due to strong gravitational lensing by foreground massive objects. 
(\S~\ref{sect:sma_results})

2. The SED of SSA22-AzTEC1 is well constrained at wavelengths $\gtrsim 3~\micron$, but drops out at a wavelength of 1~$\micron$ ($J > 25.4$ in AB, 2$\sigma$, 1$''$ aperture) or shorter. This steeply declining spectrum in NIR strongly suggests that SSA22-AzTEC1 can have a large obscuring column in front of the stellar component and/or an AGN. At the SMA position we see a significant X-ray counterpart ($\approx$20 counts), which has a hard X-ray spectrum (the photon index, $\Gamma_\mathrm{eff} = -0.34^{+0.57}_{-0.61}$). 
(\S~\ref{sect:multiw})

3. It is important to determine the redshift of SSA22-AzTEC1 to see if it is located in the protocluster, but a conventional approach using rest-frame optical-to-NIR data does not  provide a good constraint on redshift. On the other hand, photometric redshift estimates using MIR (24~$\micron$), submillimeter (860, 1100~$\micron$), and radio (20~cm) data suggest the redshift of SSA22-AzTEC1 to be $z = 3.19^{+0.26}_{-0.35}$, consistent with the protocluster redshift. 
(\S~\ref{sect:photoz})

4. We model the NIR-to-MIR SED of SSA22-AzTEC1, and find that the NIR-to-MIR SED requires large extinction ($A_V \approx$~3.4~mag) of starlight from a stellar component with $M_\mathrm{star} \sim 10^{10.9} M_{\sun}$ assuming $z \approx 3.1$. 
(\S~\ref{sect:ir_sed})

5. Modeling of the X-ray spectrum of SSA22-AzTEC1 suggests that it harbors a luminous AGN ($L_\mathrm{X} \approx 3 \times 10^{44}$~ergs~s$^{-1}$) behind a large hydrogen column ($N_\mathrm{H} \approx 1 \times 10^{24}$~cm$^{-2}$) although it is in general hard to accurately determine $L_X$ and $N_\mathrm{H}$ for AGNs with $N_\mathrm{H} \lesssim 5 \times 10^{23}$~cm$^{-2}$. These observed properties of the AGN are very similar to those of the most heavily-obscured X-ray luminous SMGs that are likely candidates for {\it protoquasars}. There are no robust candidates for protoquasars in high-$z$ protoclusters other than SSA22-AzTEC1, and it is thus possible that SSA22-AzTEC1 is the first example of a protoquasar growing at the core of a protocluster.
(\S\S~\ref{sect:smbh} and \ref{sect:protoquasar})

\medskip 

Spectroscopically determining the redshift for SSA22-AzTEC1 is particularly important to confirm the early growth of a massive black hole, as well as its host galaxy, in this high-density environment. Blind searches of CO molecular emission line(s) are necessary for spectroscopically identifying this object that is invisible in the optical and NIR. Zpectrometer \citep{Harris07} on the Green Bank Telescope, Z-Spec \citep{Naylor03} on Caltech Submillimeter Observatory, EMIR on the IRAM 30-m telescope \citep[e.g.,][]{Weiss09a}, the Redshift Search Receiver \citep{Erickson07, Chung09} on the Large Millimeter Telescope, the 32-GHz wide-band spectrometer on the 45-m telescope at Nobeyama Radio Observatory, and the Atacama Large Millimeter/submillimeter Array \citep[ALMA,][]{Wootten09, Iguchi09} will provide perhaps to achieve this.

% ==========================================================

\acknowledgments
% - acknowledge to the referee.
We are grateful to the referee for constructive suggestions which improved the presentation of this work.
% - personal 
YT thanks N.\ Kawakatu, T.\ Kodama, K.\ Kawara, T.\ Oshima for fruitful discussions. 
% JSPS, NRO
YT and BH are financially supported by the Japan Society for the Promotion of Science (JSPS) for Young Scientists. 
DMA thanks the Royal Society and a Philip Leverhulme Prize for funding. 
The Nobeyama Radio Observatory is a branch of the National Astronomical Observatory of Japan, the National Institute of Natural Sciences (NINS). 
The Submillimeter Array is a joint project between the Smithsonian Astrophysical Observatory and the Academia Sinica Institute of Astronomy and Astrophysics and is funded by the Smithsonian Institution and the Academia Sinica. 
This work is based in part on archival data obtained with the NASA Spitzer Space Telescope.

{\it Facilities:} \facility{SMA},  \facility{ASTE (AzTEC)}, \facility{VLA}, \facility{Spitzer (IRAC, MIPS)}, \facility{Subaru (Suprime-Cam, MOIRCS)}, \facility{HST (ACS)}, \facility{GALEX (NUV, FUV)}, \facility{CXO (ASIS-I)}

%%%%%%%%%%%%%
%%  - Reference -  %%
%%%%%%%%%%%%%

% ======================================================

\clearpage

% ===

\begin{deluxetable}{llccc}
\tablewidth{0pt}
\tablecaption{Multiwavelength counterparts to SSA22-AzTEC1. \label{tab1_Multiwavelength}}
\tablehead{
	\colhead{Instrument} & \colhead{Band} & \colhead{Flux density} & \colhead{Unit} & \colhead{Ref}
}
\startdata
VLA  &  20 cm  & $42 \pm 12$ & $\mu$Jy  & 1 \\
AzTEC/ASTE  & 1100 $\micron$  &  $8.2 ^{+0.8}_{-1.0}$ \tablenotemark{a} & mJy & 2 \\
SMA  &  860 $\micron$  &  $12.2 \pm 2.7$\tablenotemark{b} & mJy & 1  \\
MIPS/\textit{Spitzer}  & 24 $\micron$ &  $142 \pm 18$ & $\mu$Jy & 1 \\
IRAC/\textit{Spitzer}  & 8.0 $\micron$ &  $32.59 \pm 2.27$ & $\mu$Jy & 1 \\
 & 5.8 $\micron$  &  $22.24 \pm 1.87$\tablenotemark{c} & $\mu$Jy & 1 \\
 & 4.5 $\micron$  & $14.46 \pm 0.76$\tablenotemark{c} & $\mu$Jy & 1 \\
 & 3.6 $\micron$  & $7.80 \pm 0.50$\tablenotemark{c} & $\mu$Jy & 1 \\
MOIRCS/Subaru & 2.14 $\micron$ (\textit{K$_S$}) & $2.71 \pm 0.63$\tablenotemark{c} & $\mu$Jy & 1 \\
 & 1.64 $\micron$ (\textit{H}) & $2.13 \pm 0.71$\tablenotemark{c} & $\mu$Jy & 1  \\
 & 1.26 $\micron$ (\textit{J}) & $< 1.12$\tablenotemark{c} & $\mu$Jy & 1 \\
S-Cam/Subaru & 905~nm ($z^{\prime}$) & $< 0.080$\tablenotemark{d} & $\mu$Jy & 1 \\
 & 771~nm ($i^{\prime}$) & $< 0.042$\tablenotemark{d} & $\mu$Jy & 1 \\
 & 655~nm ($R$) & $< 0.032$\tablenotemark{d} & $\mu$Jy & 1 \\
 & 549~nm ($V$) & $< 0.035$\tablenotemark{d} & $\mu$Jy & 1 \\
 & 498~nm ($NB497$) & $< 0.050$\tablenotemark{d} & $\mu$Jy & 1 \\
 & 448~nm ($B$) & $< 0.038$\tablenotemark{d} & $\mu$Jy & 1 \\
ACIS-I/\textit{Chandra}  & 0.5--8 keV  & $2.71$ \tablenotemark{e} & $10^{-15}$~erg~cm$^{-2}$~s$^{-1}$  & 3 \\
  & 0.5--2 keV  & $9.27$ \tablenotemark{e} & $10^{-17}$~erg~cm$^{-2}$~s$^{-1}$ & 3  \\
  & 2--8 keV  & $2.66$ \tablenotemark{e} & $10^{-15}$~erg~cm$^{-2}$~s$^{-1}$ & 3 
\enddata
		\tablenotetext{a}{De-boosted flux density, i.e., flux density corrected for the flux bias due to confusion noise, with the 68\% confidence interval.}
		\tablenotetext{b}{860-$\micron$ flux density estimated from the visibility fitting.}
		\tablenotetext{c}{Aperture-corrected flux density measured using a $3\farcs 0$ aperture after the image is smoothed so that the PSF matches that of 8.0-$\micron$ image.}
		\tablenotetext{d}{$2\sigma$ upper limit measured using a $2\farcs 0$ aperture.}
		\tablenotetext{e}{Flux integrated over the ACIS-I bands.}
		\tablerefs{
			(1) This work; (2) \citet{Tamura09}; (3) \citet{Lehmer09b}.
		}
\end{deluxetable}%

% ===

\begin{deluxetable}{cccccc}
\tablecolumns{3}
\tablewidth{0pt}
\tablecaption{The properties of SSA22-AzTEC1. \label{tab2_Summary}}
\tablehead{
	\multicolumn{2}{c}{Coordinate (J2000)\tablenotemark{a} } & 	
	\colhead{$z_\mathrm{phot}$\tablenotemark{b} } &
	\colhead{$L_\mathrm{FIR}$\tablenotemark{c} } &
	\colhead{$M_\mathrm{star}$\tablenotemark{d} } &
	\colhead{$L_\mathrm{X}$\tablenotemark{e} }\\
	\cline{1-2}
	\colhead{Right ascension} & \colhead{Declination} & 
	\colhead{}  & 
	\colhead{($10^{13} L_\sun$)} &
	\colhead{($10^{10} M_\sun$)} &
	\colhead{(ergs~s$^{-1}$)}
}
\startdata
	$\mathrm{22^h17^m32 \fs 42}$ &    $+0\arcdeg 17\arcmin 44\farcs 01$ &
	$3.19^{+0.26}_{-0.35}$ &
	$1.9^{+0.4}_{-0.6}$ &
	$7.3^{+7}_{-1.7}$ &
	$\approx 3 \times 10^{44}$
\enddata
	\tablenotetext{a}{SMA 860-$\micron$ position with an uncertainty of $0 \farcs 33$.}
	\tablenotetext{b}{Photometric redshift estimate. The uncertainty is estimated from the 99\% confidence interval.}
	\tablenotetext{c}{Far-infrared luminosity. The uncertainty is estimated from the 99\% confidence interval.}
	\tablenotetext{d}{Stellar mass estimates derived from population synthesis models developed by \citet{Maraston05}. See text for the details.}
	\tablenotetext{e}{The absorption corrected luminosity in rest-frame 0.5--8~keV.}
\end{deluxetable}

% ======================================================

% ===

% ===

% ===

% ===

% ===

% ======================================================

\end{document}